\documentclass[english,prx,superscriptaddress,twocolumn]{revtex4-2}
\usepackage[T1]{fontenc}
\usepackage[latin9,utf8]{inputenc}
\usepackage{babel}
\usepackage{bm}
\usepackage{amsmath}
\usepackage{graphicx}
\usepackage{eufrak}
\usepackage{verbatim}
\usepackage{xcolor}
\usepackage{bbold}
\usepackage{float}
\usepackage{cancel}
\usepackage{xr}
\usepackage[unicode=true,pdfusetitle,bookmarks=true,bookmarksnumbered=false,bookmarksopen=false,breaklinks=false,pdfborder={0 0 0},pdfborderstyle={},backref=false,colorlinks=false]{hyperref}
\DeclareMathAlphabet{\mathpzc}{OT1}{pzc}{m}{it}

\makeatletter
\setcitestyle{numbers,square}
\makeatother

\begin{document}
\title{Quantum engineering of high harmonic generation}
\author{N. Boroumand}
\affiliation{Department of Physics, University of Ottawa, Ottawa, Ontario K1N 6N5, Canada}
\author{A. Thorpe}
\affiliation{Department of Physics, University of Ottawa, Ottawa, Ontario K1N 6N5, Canada}
\author{G. Bart}
\affiliation{Department of Physics, University of Ottawa, Ottawa, Ontario K1N 6N5, Canada}
\author{L. Wang}
\affiliation{Department of Physics, University of Ottawa, Ottawa, Ontario K1N 6N5, Canada}
\author{D. N. Purschke}
\affiliation{Joint Attosecond Science Laboratory, National Research Council of Canada and University of Ottawa, Ottawa, Ontario, K1A 0R6, Canada}
\author{G. Vampa}
\email[]{gvampa@uottawa.ca}
\affiliation{Joint Attosecond Science Laboratory, National Research Council of Canada and University of Ottawa, Ottawa, Ontario, K1A 0R6, Canada}
\author{T. Brabec}
\email[]{brabec@uottawa.ca}
\affiliation{Department of Physics, University of Ottawa, Ottawa, Ontario K1N 6N5, Canada}

\begin{abstract}
\noindent
In quantum sideband high harmonic generation (QSHHG), high harmonic generation is perturbed by a bright quantum field resulting in
harmonic sidebands, with the intent to transfer non-classical properties from the quantum perturbation to the harmonic sidebands.
So far, non-classical features have not been found in QSHHG yet. The closed form theory of QSHHG in atoms and solids
developed here answers the question under which conditions non-classical features can be realized. QSHHG results in a multi-mode
entanglement between harmonic sideband modes and perturbative quantum mode. A projective measurement on either creates a variety 
of non-classical states commonly used in quantum information science. This opens a pathway towards quantum engineering high 
harmonic generation as a short wavelength source for quantum information science. 
\end{abstract}
\maketitle

\section{Introduction}

\noindent
High harmonic generation (HHG) is a coherent source for XUV (extreme ultraviolet) generation and presents the key process for many 
spectroscopic applications in attosecond science \cite{krausz2009attosecond}. It is successfully described within the 
semi-classical approximation, whereby the atomic dynamics are described quantum mechanically in conjunction with a classical 
model of the electromagnetic field \cite{lewenstein1994theory}. This approach allows a simple interpretation of HHG in terms of a 
three-step process, consisting of ionization, free electron quiver excursion driven by the laser field, and emission of high 
harmonic radiation during recombination with the parent ion \cite{corkum1993plasma}. Over the past few decades, significant 
progress in understanding and utilization of HHG in both gaseous and solid media has been made possible via the use of the semi-classical approximation \cite{goulielmakis2022high, ghimire2011observation}.

Only recently has HHG been analyzed from a quantum optical perspective. By describing the harmonic emission process with quantum 
optical operators and modeling the intense laser field classically, it has been demonstrated that the photon statistics of high
harmonic (HH) radiation can be modified under certain conditions \cite{gonoskov2016quantum,gorlach2020quantum}, such as by
accounting for the electron dynamics between bound states. HHG has also been described in two-level atoms using quantum optical
models \cite{foldi2021describing,gombkotHo2021quantum} and has been shown both theoretically and experimentally to induce 
non-classical Schr\"odinger cat states in the intense laser modes 
\cite{tsatrafyllis2019quantum,rivera2021new,lewenstein2021generation,stammer2023quantum}. Additional intriguing predictions from
theory include light-electron entanglement during above threshold ionization \cite{rivera2022light}, entangled x-ray photon pair
generation through HHG from pair production \cite{sloan2023entangling}, and entanglement between harmonics in intense laser driven 
atoms, when a resonance lies close to a harmonic \cite{yi2025generation}. In the limit of strong field driving, when the atomic
ground state gets depleted, the quantum state of the driving field and harmonics become entangled and squeezed
\cite{stammer2024entanglement}. 

Furthermore, theoretical and experimental investigations have started to look into the modification of strong field physics in the 
case where the intense laser field is replaced by a bright squeezed vacuum (BSV) beam. Full quantum optical calculations have 
predicted substantial deviations from the semi-classical three-step model, the most notable of which are harmonics beyond the 
semi-classical cutoff, higher damage thresholds, and the photon statistics of the BSV influencing the emitted electrons 
\cite{even2023photon,even2024motion,rasputnyi2024high,heimerl2024multiphoton}. 

Developing quantum sources at shorter wavelengths is a desirable goal, because of the higher information density and lower noise of 
detectors. The required intensity for extending HHG into the XUV (extreme ultraviolet) range presents a challenge, if the only 
driving field is intended to be quantum mechanical. Ideally, for applications in quantum information science \cite{rivera2022light}
one desires a coherent process that can be scaled to the XUV, in which quantum properties can be engineered at will. These have 
been the guiding principles behind a recent experiment: perturbing regular HHG with a BSV to produce quantum sideband high 
harmonics (QSHH) that exhibit super-Poissonian photon-bunching statistics \cite{lemieux2024photon}. 

The goal of this work is to develop the theory of quantum sideband high harmonic generation (QSHHG) in atoms and solids and to 
identify methods by which to transfer quantum properties from the perturbative BSV onto the harmonic sidebands. The theoretical 
framework is a quantum generalization of the semi-classical Lewenstein model of HHG \cite{lewenstein1994theory} and yields 
closed-form solutions for the HHG and QSHHG wavefunctions. Knowledge of the wavefunction enables identification of the quantum 
properties of QSHHG. The additional photons absorbed and emitted from the quantum perturbation, such as a BSV, create entanglement 
between individual harmonic sidebands and between the harmonic sidebands and the BSV. We show how this entanglement can be 
harnessed to create a variety of non-classical states commonly used in quantum information science, such as high purity single
photon states, Schr\"odinger cat states, and photon added squeezed vacuum states. In this way, quantum properties of the BSV can be 
transferred onto QSHHG, opening a path towards engineering the quantum properties of ultrashort XUV high harmonics. While we 
primarily focus on single-mode properties, a qualitative discussion of two- and multi-mode entanglement is given in the following 
text. A more quantitative analysis is subject to future research.

In addition to revealing quantum properties of QSHHG, the theory offers an order of magnitude predictive power regarding the number 
of photons in the harmonics and sidebands and compares very favorably with experiments \cite{lemieux2024photon}. This facilitates 
the optimization of QSHHG, thereby relaxing the requirements on the BSV power. 

\section{Theory summary}

\noindent
We use a strong field quantum optical model that generalizes the semi-classical approach of Lewenstein 
\cite{corkum1993plasma, lewenstein1994theory}. Although the derivation is quite general, results and conclusions are 
derived for a BSV quantum perturbation at twice the frequency of the classical driving field which results in even harmonic sidebands. For 
a detailed derivation see the supplement \cite{supp}. In the quantum optical description, HHG is modeled as a one photon 
process: $N$ classical photons are emitted as a single high harmonic (HH) photon. QSHHG is, to lowest order, a two-photon 
process, wherein the emission of a harmonic photon is perturbed by the emission or absorption of a quantum photon, resulting 
in the effective emission of a QSHH photon. The international system of (SI) units is used, unless otherwise noted. 

\subsection{Wavefunction}

\noindent
HHG and QSHHG are described by the photon wavefunction 
\begin{align}
& \left \vert \varphi_0(t) \right \rangle \! \approx \! \left \vert \varphi_{h}(t) \right \rangle \left \vert \varphi_{m}(t) \right \rangle \!=\! 
\hat{D}_{h} \hat{S}_m \left \vert \varphi_0(t_0) \right \rangle \mathrm{}
\label{hhqc1}
\end{align}
with $\vert \varphi_0(t_0) \rangle$ the wavefunction at initial time $t_0$. We assume that HHG and QSHHG take place in different modes $\kappa$. The wavefunction can therefore be written as an independent product of these two processes. For example, coupling between QSHHG and HHG occurs for three-photon processes, when the frequency of the perturbing quantum field is twice that of the 
classical one \cite{bertrand2011ultrahigh, purschke2023microscopic}. It can also occur in a first order process when the perturbing and driving fields have the same frequency. Both cases are 
subject to future research. A generalization of our approach is outlined in the supplement \cite{supp}. HHG and QSHHG are determined in the limit $\left \vert \varphi_{h,m} \right \rangle = 
\left \vert \varphi_{h,m}(t \rightarrow \infty) \right \rangle$. 

The HHG wavefunction $\left \vert \varphi_{h} \right \rangle$ is generated by the displacement operator 
\begin{align}
\hat{D}_{h} = \exp \left( \sum_{\kappa} h_{\kappa} \hat{a}^{\dagger}_{\kappa} - h^*_{\kappa} \hat{a}_{\kappa} \right) \mathrm{,}
\label{hhq}
\end{align}
where $\kappa \equiv (\mathbf{k}s)$ is a multi-index with photon wavevector $\mathbf{k}$, polarization index $s=1,2$. The HHG 
coefficient $h_{\kappa}$ is defined in the next section; it contains a sum over an ensemble of emitters at positions $\mathbf{x}_j$. The position dependence is not written explicitly. 

The QSHHG wavefunction $\left \vert \varphi_{m}(t) \right \rangle$ is generated by a mixed-mode squeezed vacuum state operator between harmonic modes $\kappa$ and perturbative quantum modes $q$, 
\begin{align}
\hat{S}_{m}  = \exp \left( \sum_{\kappa} \left( f_{\kappa} \hat{a}^{\dagger}_{q} + g_{\kappa} \hat{a}_{q} \right) \hat{a}^{\dagger}_{\kappa} - \left( g^{*}_{\kappa} \hat{a}^{\dagger}_{q} 
+ f^{*}_{\kappa} \hat{a}_{q} \right) \hat{a}_{\kappa} \right) \mathrm{.}
\label{qshhq}
\end{align}
Our analysis is confined to a single perturbative mode $q$. The coefficients $f_{\kappa}, g_{\kappa}$ represent HHG in the presence of emission and absorption of an additional perturbative quantum 
photon, respectively. 
In perturbative nonlinear optics, the former is referred to as difference (DFG) and the latter as sum frequency (SFG) generation.
QSHHG coefficients $f_{\kappa}, g_{\kappa}$ are defined in the next subsection. 

In order to fully define Eq. (\ref{hhqc1}), a particular initial state needs to be chosen; $\vert \varphi_0(t_0) \rangle = \vert v \xi_q \rangle$ consists of a multi-mode vacuum state,  
$\vert v \rangle = \prod_{\kappa} \vert v_{\kappa} \rangle$, for the harmonics, and a single squeezed vacuum state for the perturbative mode, 
\begin{align}
\left \vert \xi_q \right \rangle = \hat{S}_q \left \vert v_q \right \rangle = e^{\frac{1}{2}(\xi \hat{a}_q^2 -\xi^* 
\hat{a}_q^{\dagger 2})} \left \vert v_q \right \rangle
\label{squvac}
\end{align}
with $\xi = r e^{i \theta}$. Normal ordering of $\left \vert \varphi_{m} \right \rangle \!=\! \hat{S}_m \left \vert v \xi_{q} \right \rangle$ yields \cite{supp}
\begin{align}
\left \vert \varphi_{m} \right \rangle \!=\! \frac{\overline{\frak{N}}}{\sqrt{\cosh(r)}} \exp \left( \sum_{\kappa} Z_{\kappa} \hat{a}_{\kappa}^{\dagger} 
\hat{a}_{q}^{\dagger} \right) \exp \left(- \overline{\beta} \hat{a}_q^{\dagger 2} \right) \left \vert v v_q \right \rangle 
\label{nord}
\end{align}
with $\beta = (1/2) \tanh(r) e^{-i \theta}$, and $\overline{\beta} = \beta / (1 + \sum_{\kappa} \vert Z_{\kappa} \vert^2)$. Further, 
\begin{align}
Z_{\kappa} & = f_{\kappa} - g_{\kappa} \tanh(r) e^{i \theta} \nonumber \\
\overline{\frak{N}} & = \frac{1}{\sqrt{1 + \sum_{\kappa} \vert Z_{\kappa} \vert^2} } \mathrm{.}
\label{nordcoef}
\end{align}
Equation (\ref{nord}) is a multi-mode harmonics wavefunction in the basis of electromagnetic plane wave modes, and can be used to calculate two or more correlated mode properties. The normal-ordered wavefunction has been derived in the limit of an intense squeezed vacuum beam with $r > 1$ and $\vert Z_{\kappa} \vert^2 \ll 1$.

\subsection{HHG and QSHHG coefficients for atomic and molecular gases}
\label{coefficients}

\noindent
The HHG coefficient of a single atom $j$ is given by \cite{supp}
\begin{align}
h_{\kappa} & = - e^{-i \mathbf{k} \mathbf{x}_j} \!\! \int_{-\infty}^{\infty} \!\!\!\! dt e^{i \omega_k t } H_{k}(t) =  
\tilde{H}_{k} e^{-i \mathbf{k} \mathbf{x}_j} \label{hhg} \\
H_{k} & = \frac{\vert e \vert E_v}{\hbar} \Biggl\{ \mathbf{e}_{\kappa} \mathbf{x}(t)   
+ \int \!\! d^3\mkern-2mu p \, \overline{\sigma}_{\kappa}(t) \left( \Gamma_{\mathbf{p}}(t) + \mathrm{c.c} \right) \Biggr\} \mathrm{}
\nonumber 
\end{align}
with
\begin{align}
& \mathbf{x}(t) = \int \!\! d^3\mkern-2mu p \, \mathbf{d}^*\!(\mathbf{p}_t) \, b_{\mathbf{p}}(t) + \mathrm{c.c.} \label{lewdip} \\ 
& b_{\mathbf{p}}(t) = \int_{t_0}^{t} dt^{\prime} \, \Omega(t^{\prime}) \exp \left( i S(t^{\prime}, t) - \xi(t-t^{\prime})  \right)
\mathrm{.}
\nonumber 
\end{align}
The polarization of HHG is assumed to be parallel to the laser pulse; additionally, $H_{k}$ does not depend on the direction of the wavevector. Both facts are reflected in the change of the lower index from 
$\kappa$ to $k = \omega_k/c$. Note that $H_{k}$ depends on the position $\mathbf{x}_j$ of the atom via the space dependence of the laser field, which is not explicitly stated. Frequency and polarization of the harmonic are given by $\omega_k$ and $\mathbf{e}_{\boldsymbol{\kappa}}$, 
$\mathbf{d}(\mathbf{p}) = \langle \mathbf{p} \vert \mathbf{x} \vert 0 \rangle = d_0 \mathbf{p} / (\mathbf{p}^2/(2m) + E_0)^3 $ 
is the transition dipole moment between ground $\vert 0 \rangle$ and continuum plane wave state $\vert \mathbf{p} \rangle$ with 
$\mathbf{p}$ in the canonical momentum frame \cite{lewenstein1994theory}, $E_0$ is the binding energy, and $d_0 = 3.37 \times 10^{-4} \, [\mathrm{kg^{1/2} m^{9/2} s^{-7/2}}]$. Further, $\Omega(t) = (\vert e \vert /\hbar)
\mathbf{d}(\mathbf{p}_t) \boldsymbol{\mathcal{F}}(t)$ is a generalized Rabi frequency, $\boldsymbol{\mathcal{F}}$ represents the classical intense laser field, $\mathbf{p}_{\mkern-1.5mu t} = 
\mathbf{p} + |e| \boldsymbol{\mathcal{A}}(t)$ is defined in the moving momentum frame, and $- \partial_t \boldsymbol{\mathcal{A}}(t) = \boldsymbol{\mathcal{F}}(t)$ defines the vector potential. 
Note that the electric field consists of a classical part $\boldsymbol{\mathcal{F}}$ with frequency $\omega_0$ and field strength $F_0$, and of a quantum part $\hat{\mathbf{F}}$ which accounts for 
the emission of harmonic photons in Eqs. (\ref{hhq}) and (\ref{qshhq}). Moreover, $E_v = \sqrt{\hbar \omega_k / 2 \varepsilon_0 V}$
is the vacuum electric field, $V$ the quantization volume, and $\varepsilon_0$ the vacuum permittivity. We use a filter 
$\xi(t-t^{\prime})$ 
that leaves HHG from the first recollision unchanged and extinguishes all higher returns \cite{supp}. The dipole $\mathbf{x}(t)$ corresponds with the semiclassical Lewenstein dipole defined in Eq. 
(6) of Ref. \cite{lewenstein1994theory} with 
\begin{align}
S(t) = \frac{1}{\hbar} \int_{t_0}^{t} \! \left( \frac{1}{2m} \mathbf{p}^2_{\mkern-1.5mu \tau} + E_0 \right) d\tau \mathrm{}
\label{action}
\end{align}
the classical action, and $ S(t^{\prime}, t) = S(t) - S(t^{\prime})$. Further, 
\begin{align}
\Gamma_{\mathbf{p}}(t) = \Omega^*(t)  b_{\mathbf{p}}(t) 
\label{Gamma}
\end{align}
is related to the optical field ionization rate $2 \mathrm{Re}[\gamma]$ with $\gamma(t) = \int d^3\mkern-1.5mu p \, \Gamma_{\! \mathbf{p}}(t)$, see Eq. (52) of \cite{lewenstein1994theory}. 
Finally, during ionization an electron is promoted into a laser driven continuum state, dressed with a displacement operator \cite{supp} with coefficient 
\begin{align}
& \sigma_{\kappa}(t) = \! \frac{\vert e \vert E_{v}}{\hbar} \overline{\sigma}_{\kappa}(t) e^{i \omega_{k} t } \nonumber \\ 
& \overline{\sigma}_{\kappa}(t) = \! 
- \frac{i}{\omega_k} \! \int_{t_0}^{t} dt^{\prime}  \left( \mathbf{e}_{\kappa} \mathbf{v}_{t^{\prime}} \right)  
e^{-i \omega_{k} (t - t^{\prime})} \mathrm{.}
\label{sigma}
\end{align}
Here, $\mathbf{v} = \mathbf{p}/m$ is the electron velocity. 
HHG, as described by $H_k$ in Eq. (\ref{hhg}), contains two contributions. The first term represents HHG via ionization, continuum evolution, and recollision \cite{corkum1993plasma,lewenstein1994theory}.
The second term describes HHG via the ionization nonlinearity \cite{brunel1990harmonic}. 

The QSHHG coefficients $f_{\kappa}, g_{\kappa}$ for a single atom $j$ are given by 
\begin{align}
f_{\kappa} & \!=\! e^{-i (\mathbf{k} + \mathbf{q}) \mathbf{x}_j} \!\!\! \int_{-\infty}^{\infty} \!\!\!\!\! dt e^{i (\omega_{k}+\omega_{q}) t} F_{k}(t) 
\! = \! \tilde{F}_{k} e^{-i (\mathbf{k} + \mathbf{q}) \mathbf{x}_j} \nonumber \\
F_{k} & \!=\! \left(\! \frac{\vert e \vert E_{v}}{\hbar} \! \right)^{\!2} \!\!\! \int \!\! d^3\mkern-2mu p \, \overline{\sigma}_{q}(t) \biggl\{  \mathbf{e}_{\kappa} \mathbf{x}_{\mathbf{p}}(t) 
\!+\! i \overline{\sigma}_{\kappa}(t) \mathrm{Im}\left[ \Gamma_{\mathbf{p}}(t) \right] \biggr\} 
\label{f2ph} 
\end{align}
and  
\begin{align}
g_{\kappa} & \!=\! e^{-i (\mathbf{k} - \mathbf{q}) \mathbf{x}_j} \!\!\! \int_{-\infty}^{\infty} \!\!\!\!\! dt e^{i (\omega_{k}-\omega_{q}) t} G_{k}(t) 
\! = \! \tilde{G}_{k} e^{-i (\mathbf{k} - \mathbf{q}) \mathbf{x}_j} \nonumber \\
G_{k} & \!=\! \left(\! \frac{\vert e \vert E_{v}}{\hbar} \! \right)^{\!2} \!\!\! \int \!\! d^3\mkern-2mu p \, \overline{\sigma}^*_{q}(t) 
\biggl\{\mathbf{e}_{\kappa} \mathbf{x}_{\mathbf{p}}(t) \!+\! i \overline{\sigma}_{\kappa}(t) \mathrm{Im}\left[ \Gamma_{\mathbf{p}}(t) \right] \biggr\} \mathrm{,}
\label{g2ph} 
\end{align}
where $\omega_q$ and $\mathbf{q}$ are frequency and wavevector of the quantum light mode $q$, and 
\begin{align}
\mathbf{x}_{\mathbf{p}}(t) = \mathbf{d}^* \!(\mathbf{p}_t) \, b_{\mathbf{p}}(t) - \mathrm{c.c.} 
\label{x-}
\end{align}
is the imaginary part of the transition dipole for a given $\mathbf{p}$. Again, the index of $F,G$ has been changed from $\kappa$ to $k$. Both, conventional HHG and QSHHG coefficients are very similar to the semi-classical coefficients \cite{purschke2023microscopic}, except for powers of the quantum vacuum field term $\vert e \vert E_{v} / \hbar$. In the quasi-classical approach 
the coefficient $\overline{\sigma}_{q}(t)$ is within the time integral of $\mathbf{x}_{\mathbf p}(t)$. The same is initially the case for the quantum optical equations. However, in order to 
obtain a unitary QSHHG operator, $\overline{\sigma}_{q}$ needs to be pulled out of the inner time integral by integration by parts. The remaining non-unitary term is small and can be neglected, 
see the discussion between Eqs. (S14) and (S15) in section I.A of the supplement \cite{supp}. 

\subsection{HHG and QSHHG coefficients for solids}
\label{transition}

\noindent
The results for atomic and molecular gases can be easily translated into HHG in two-band solids by replacing 
\begin{subequations}
\label{replace}
\begin{align}
& \mathbf{p} \rightarrow \mathbf{k}, \,\,\,\, \mathbf{k} \in \mathrm{BZ} \label{replace1} \\
& E_0 + \frac{\mathbf{p}^2}{2m} \rightarrow \varepsilon(\mathbf{k}) \approx E_g + \frac{\mathbf{p}^2}{2m_{*}} = \varepsilon(\mathbf{p}) \label{replace2} \\
& \frac{\mathbf{p}}{m} \rightarrow \frac{1}{\hbar} \nabla_{\mathbf{k}} \varepsilon = \mathbf{v}(\mathbf{k}) \approx \frac{\mathbf{p}}{m_{*}} \label{replace3} \\
& \mathbf{d}(\mathbf{p}) \rightarrow \mathbf{d}(\mathbf{k}) = -i \langle u_c(\mathbf{k}) \vert \nabla_{\mathbf{k}} \vert u_v(\mathbf{k}) \rangle \approx 
\frac{d_0 E_g}{\varepsilon(\mathbf{p})}
\label{replace4}
\end{align}
\end{subequations}
where $\mathbf{k}$ represents the crystal momentum defined in the first Brillouin zone (BZ), and $\varepsilon(\mathbf{k})$ and $\mathbf{v}(\mathbf{k})$ represent relative band gap and band velocity, 
i.e. the difference between conduction and valence bands. The minimum band gap is $E_g = E_0$, $m_{*}$ is the effective electron mass at the band gap minimum defined by $\partial_{k_i} 
\partial_{k_j} \varepsilon \vert_{\mathbf{k}=0} = \hbar^2 / m_{*}$. For simplicity, we have replaced the inverse effective mass tensor by a scalar quantity. Finally, the atomic transition 
dipole needs to be replaced with the dipole moment between valence and conduction bands. The last term is the approximate dipole moment in $\mathbf{k}\mathbf{p}$ perturbation theory 
\cite{vampa2014theoretical}. 
Within the effective mass approximation, the atomic equations remain applicable to solids and are represented by the last terms in Eqs. (\ref{replace1})-(\ref{replace4}). 

\subsection{Phase matching}

\noindent
In order to compare and characterize HHG and QSHHG, operator expectation values with regard to the macroscopic wavefunction need to be known. Here, we focus on $\hat{n} = \sum_{\kappa} 
\hat{n}_{\kappa}$, 
\begin{align}
& \langle \varphi_h \vert \hat{n} \vert \varphi_h \rangle = \sum_{\kappa} \! \left( \sum_j h^*_{\kappa} \right) \!\! 
\left( \sum_j h_{\kappa} \right) \nonumber \\
& = \frac{V N_0^{2}}{(2\pi)^3} \! \int \!\! d^3\mkern-1.5mu k \left( \int \!\! d^3\mkern-1.5mu x \, h^*_{\kappa} \! \right) \left( \int \!\! d^3\mkern-1.5mu x \, h_{\kappa} \! \right) 
\mathrm{.}
\label{expecthh}
\end{align}
In the continuum limit, the sum over atoms $\sum_i \rightarrow N_0 \int d^3\mkern-1.5mu x$, with $N_0$ representing the number density of the material and $\sum_{\kappa} \rightarrow V/(2\pi)^3 \int 
d^3\mkern-1.5mu k$, with $V$ as the quantization volume. Polarization and wavevector of intense laser, quantum field, HHG, and QSHHG are fixed along $x$ and $z$, respectively. The space integrals and the transverse $k_x$- and $k_y$-integrals can be worked out approximately (see \cite{supp}). At this point, the only remaining integral is the one over $dk_z \approx dk = d \omega_k / c$. The $d\omega_k$ integral drops out upon examination of the differential expectation value of the number operator, 
\begin{align}
\frac{d \langle \hat{n} \rangle}{d \omega_k} & = \frac{d}{d \omega_k}\langle \varphi_h \vert \hat{n} \vert \varphi_h \rangle = c_{k}^2 \, \vert \tilde{H}_{k}(\omega_k) \vert^2 
\nonumber \\
c_{k}^2 & = \frac{ \left( N_0 \mathrm{w}_{k} l_i \right)^2}{2c} V \mathrm{,} 
\label{expecthh1}
\end{align}
where $\mathrm{w}_{k} = \mathrm{w}(\omega_{k})$ represents the transverse $1/e^2$ radius of the HH mode with frequency $\omega_k$. Here, $\tilde{H}_{k}=\tilde{H}_{k}(\mathbf{x}=0)$, since the space dependence has 
been integrated over, and the interaction length is $l_i$. HHG scales as $l_i^2$, as long as $l_i$ is shorter than the dephasing length. Finally, the quantization volume cancels out due to the fact that the single atom response $\tilde{H}_{k} \propto 1/\sqrt{V}$. 

We are mainly interested in the number of photons emitted in all spatial modes with frequencies in the band  $(N-\frac{1}{2})\omega_0 \le \omega_k \le (N+\frac{1}{2})\omega_0$ which is given by 
\begin{align}
\langle \hat{n} \rangle_{\!N} = c_{k}^2 \int_{(N-\frac{1}{2})\omega_0}^{(N+\frac{1}{2})\omega_0} \!\!\!\! d \omega_k 
\vert \tilde{H}_{k}(\omega_k) \vert^2 = \vert h_{N} \vert^2
\mathrm{,}
\label{expecthh2}
\end{align}
with $h_{N}$ being a dimensionless quantity. 

Similarly, the number of QSHHG photons emitted into one harmonic interval $\omega_0$ about harmonic order $N$ is found to be \cite{supp}
\begin{align}
\langle  \hat{n} \rangle_{\!N} \!=\! c_q^2 \cosh^2(r) \int_{(N-\frac{1}{2})\omega_0}^{(N+\frac{1}{2})\omega_0} \!\!\!\! d \omega_k \left \vert \zeta_{k} \right \vert^2 
= \cosh^2 (r) \left \vert \zeta_{N} \right \vert^2 
\mathrm{}
\label{nqsom0}
\end{align}
with 
\begin{align}
& \vert \zeta_{k} \vert^2 = \left \vert \tilde{F}_{k} - \tilde{G}_{k} \tanh(r) e^{i \theta} \right \vert^2 \mathrm{,} \nonumber \\
& c_q^2 = \frac{ \left( N_0 \mathrm{w}_{k} l_i \right)^2}{2c} V^2 \frac{(\Delta q)^3}{(2 \pi)^3} \mathrm{,}
\label{cqsq}
\end{align}
and $(\Delta q)^3$ is the mode volume of the perturbative quantum field. It should be noted that we have approximated the temporally and spatially finite BSV field by a plane wave. 
This is being corrected for by replacing the plane wave mode volume by the experimentally measured BSV mode volume \cite{lemieux2024photon}. The parameter $\zeta_k$ emerges from $Z_{\kappa}$ in Eq. 
(\ref{nordcoef}) after performing the phase matching integrals \cite{supp}. The quantization volume again drops out. Similar to the case for HHG, $\zeta_{N} = \zeta_{N}(\mathbf{x}=0)$. 

\subsection{Effective QSHH mode}
\label{QSHHmode}

\noindent
In experiments \cite{lemieux2024photon}, the photons contained in one QSHH are measured and \textit{not} in the individual plane wave modes. To that end, the connection between calculations and experiment is greatly facilitated by 
introducing an effective mode operator \cite{rohde2007spectral}
\begin{align}
\hat{a}_N = \frac{1}{\vert \zeta_{N} \vert}\sum_{\kappa \in N} Z^*_{\kappa} \hat{a}_{\kappa}
\label{1effmo}
\end{align}
that encompasses all plane wave modes of a quantum sideband. This operator fulfills the usual harmonic oscillator commutation relations $[\hat{a}_N,\hat{a}^{\dagger}_M] = \delta_{NM}$. 
The vacuum state of a QSHH mode is $\prod_{\kappa \in N} \vert v_{\kappa} \rangle = \vert v \rangle_N$, so that a number state
\begin{align}
\vert n \rangle_{N} = \frac{1}{\sqrt{n!}} \left( \hat{a}_N^{\dagger} \right)^n \vert v_{N} \rangle \mathrm{,}
\label{1effmns}
\end{align}
corresponds to a sum over all combinations that have $n$ photons in the plane wave modes of the quantum sideband $N$. With these definitions the wavefunction (\ref{nord}) becomes 
\begin{align}
& \left \vert \varphi_m \right \rangle \!=\! \prod_{N} \frac{\frak{N}_{N}}{\sqrt{\cosh (r)}} \exp \! \left( \vert \zeta_{N} \vert \hat{a}_{N}^{\dagger} \hat{a}_{q}^{\dagger} \right) 
\exp \! \left(- \beta_{N} \hat{a}_q^{\dagger 2} \right) \left \vert v_{N} v_q \right \rangle \mathrm{,} \nonumber \\
& \beta_{N} = \frac{\beta}{ 1+ \sum_{\kappa \in N} \vert Z_{\kappa} \vert^2} = \frac{\beta}{ 1 + \vert \zeta_{N} \vert^2} \mathrm{,} \nonumber \\
& \frak{N}_{N} = \frac{1}{\sqrt{1+ \sum_{\kappa \in N} \vert Z_{\kappa} \vert^2}} = \frac{1}{\sqrt{1 + \vert \zeta_{N} \vert^2}} \mathrm{.}
\label{nordeff}
\end{align}
This wavefunction will be used throughout the remaining text.

\section{Results}
\label{results}

\noindent
The following section begins with the calculation of the macroscopic photon numbers emitted by HHG and QSHHG and an associated discussion. This is followed by plotting the two-mode distribution 
function, which resembles a two-mode squeezed state and is entangled. Some entanglement features are briefly discussed. The rest of the paper 
focuses mainly on single-mode properties. From the two-mode distribution function, the QSHHG distribution function is calculated. This reveals why 
the non-classical properties of the BSV perturbation are not carried over onto the QSHH state. Finally, projective measurements are discussed,
where the photon number of the QSHH ($N$) or perturbative ($q$) mode is measured, resulting in a wavefunction in the other mode which carries interesting non-classical properties, such as squeezing and a Wigner function with negative values.

\subsection{QSHHG: gases versus solids}
\label{gasvsol}

\begin{figure}[h]
\centering
\hspace*{-0.5cm}
\includegraphics[width=9.8cm]{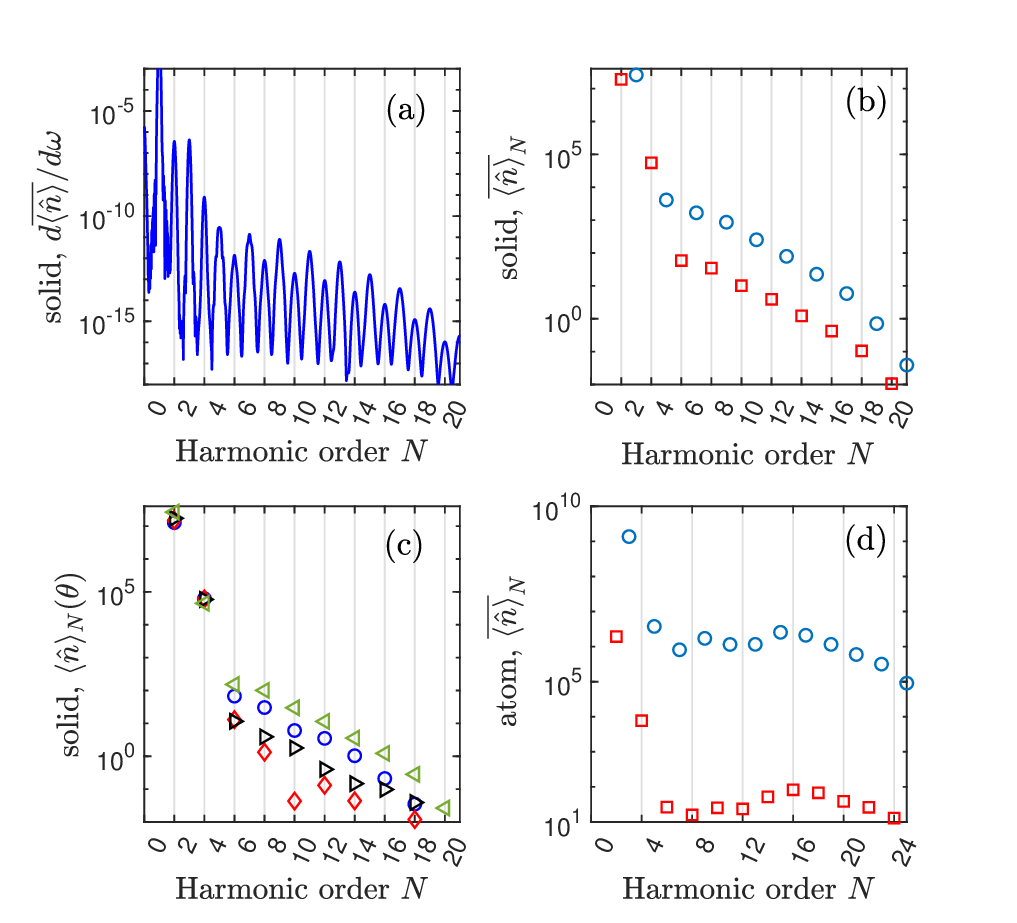}
\caption{(a-c) HHG (odd) and QSHHG (even) spectrum in ZnO versus harmonic order N; (a) differential number of HH photons 
$d\overline{\langle \hat{n} \rangle} / d \omega$; (b) number of photons in the effective mode 
$\overline{\langle \hat{n} \rangle} _{N}$; HHG (blue circles), QSHHG (red circles). (a)-(b) average over squeezed vacuum phase 
$\theta$. (c) $\langle \hat{n} \rangle_{N}(\theta)$: blue circles, red diamonds, black right-pointing triangles, green 
left-pointing triangles correspond to $\theta = 0, \pi/4, 3\pi/8, 7\pi/4$, respectively. 
(d) $\overline{\langle \hat{n} \rangle}_{N}$ averaged over $\theta$ for an atomic gas; HHG (blue circles), QSHHG (red squares). 
For parameters see text.} 
\label{fig1}
\end{figure}

\noindent
In Fig. \!\ref{fig1} QSHHG in solids (a-c) and in atoms (d) is compared. We utilize Eq. (\ref{replace}) to model the ZnO band
structure. The parameters used are $m_* \approx 0.25 m $, $E_g = 3.4$ eV, and $d_0 \!=\! 7 \times 10^{25} \, 
\left[\mathrm{{s^2/(kg^3 \cdot m)}}\right]^{1/2}$, the last of which is converted from $d_0 \!=\! 4$ atomic units in Ref. 
\cite{vampa2014theoretical}. The laser parameters are those which were used in recent experiments \cite{lemieux2024photon}: 
wavelength $\lambda_0 = 3.2$ \textmu m and peak electric field strength $F_0 = 1.3 \times 10^{9}$ V/m, corresponding to peak
intensity $I_0 = 5 \times 10^{11}$ W/cm$^2$. The electric field consists of a sine-carrier and a Gaussian envelope with pulse 
durations $\tau=6T_0$, where oscillation period $T_0 = 2\pi/\omega_0 \approx 10 $fs. The macroscopic propagation parameters 
are $N_0 = 4 \times 10^{22}$ cm$^{-3}$, $\mathrm{w}_k = 40$ \textmu m (assumed to be approximately half of the laser beam 
width due to the HHG nonlinearity), and $l_i = 5$ nm as determined by the absorption length. This results in the macroscopic 
HHG factor $(N_0 \mathrm{w}_k l_i)^2/(2c) \approx 10^{23}$ s/m$^{3}$; $\omega_q = 2 \omega_0$ is used for the squeezed vacuum 
field. From the pulse energy of $10$ nJ, \cite{lemieux2024photon} one obtains $r=13.6$ and $\sinh^2(r) \approx \cosh^2(r) = 
10^{11}$ via the relation $\hbar \omega_q \langle \hat{n}_q \rangle = \hbar \omega_q \sinh^2 r = 10^{-8}$ J. The mode volume 
of the perturbative quantum beam must also be known for macroscopic QSHHG . In accordance with experiments
\cite{lemieux2024photon}, we choose a transverse width $\mathrm{w}_q = 100$ \textmu m and bandwidth $\Delta \lambda_q = 50$ nm. 
Inserting these values into the expression for the mode volume gives $(2 \pi / \Delta q)^3 = (2 \pi)^3
\Delta \lambda_q / (\mathrm{w}_q \lambda_q)^2 = 2\times 10^{12}$ m$^3$. 

The atomic gas parameters for Fig. \ref{fig1}(d) are: hydrogen atom, $E_0 = 13.6$ eV, for dipole moment see subsection \ref{coefficients}; laser: $\lambda_0 = 800$ nm, $I_0 = 10^{14}$ W/cm$^2$, 
$\tau = 6T_0 \approx 15$ fs. The macroscopic propagation parameters are: gas nozzle, $N_0 = 10^{18}$ cm$^{-3}$, $\mathrm{w}_k = 100$ \textmu m, and $l_i = 250$ \textmu m, which result again in 
$(N_0 \mathrm{w}_k l_i)^2/(2c) \approx 10^{23}$ s/m$^{3}$. The squeezed vacuum parameters are $\omega_q = 2 \omega_0$, and the rest is the same as in Fig. \ref{fig1}(a-c). 

The differential number of HH photons averaged over the BSV angle $\theta$, $d \overline{\langle \hat{n} \rangle} / d\omega $, is plotted for ZnO in Fig. \ref{fig1}(a); (b) shows number of photons in one effective harmonic mode $N$ averaged over $\theta$, 
$\overline{\langle \hat{n}} \rangle_{N}$. The parameters are similar to recent experiments \cite{lemieux2024photon} and agree well with the ratio of HHG to QSHHG which is roughly an order of 
magnitude. Figure \ref{fig1}(c) reveals a sensitive dependence of QSHHG on $\theta$; 
variation of $\theta$ changes QSHHG by up to 3 orders of magnitude. The modulation is not only a feature of a quantum perturbation and is also found in coherent control experiments with 
bi-chromatic coherent fields \cite{dudovich2006measuring, vampa2015linking}. 

Figure \ref{fig1}(d) shows $\langle \hat{n} \rangle_{N}$ for gaseous atomic hydrogen to be compared with (b) for ZnO. In contrast to (b), the difference between QSHHG and HHG in the gas is six orders of
magnitude. This can be understood by looking at the different scaling of HHG and QSHHG in Eqs. (\ref{hhg}) and (\ref{f2ph}), (\ref{g2ph}). The main difference comes from the additional 
$\overline{\sigma}_q$ dependence outside the curled brackets in Eqs. (\ref{f2ph}) and (\ref{g2ph}). By denoting the ratio of QSHHG and HHG as $R$ -- a measure of the susceptibility to the quantum perturbation -- we see that 
\begin{align}
R \propto \vert \sigma_q \vert^2 \propto \frac{I_0}{\omega_0^2} \frac{1}{\omega_q^3 m_*^2} \propto \frac{I_0}{8\omega_0^5 m_*^2} \mathrm{.}
\label{rel_diff}
\end{align}
The first factor originates from $\mathbf{p}^2_t$. Note that $E_v^2 \propto \omega_q$, so that the frequency scaling of the second factor is $1/\omega_q^3$ and not 
$1/\omega_q^4$, as semiclassical theory would predict. The last proportionality stems from $\omega_q = 2\omega_0$. 

The parameters corresponding to Fig. \ref{fig1} include a factor of 4 reduction of both laser frequency and effective mass in the case of the solid.
This contributes an overall factor of $4^7=16384$, which is uncompensated by the $200$ times increase of intensity in the gas, and results in $R$ increasing by a factor of $\approx 100$. The difference of the solid:gas ratio in Fig. \ref{fig1} is slightly less than 1000. This demonstrates that the simple estimate (\ref{rel_diff}) presents a lower limit to the scaling of the solid:gas ratio. 

Although solids are more susceptible to the quantum perturbation, the higher efficiency of HHG in atoms results in similar $\hat{n}_N$ for QSHHG in solids and atoms in Figs. \ref{fig1}(b) and (d). 
What is unique to solids is that the factor $R$, and thus the conversion efficiency, can be further increased by selecting materials with low effective mass. Choosing lasers with longer wavelengths is
equally favorable for atoms and solids. For example, Bi related materials have $m_{*}/m \approx 0.002$ \cite{reshak2014density} with $m$ being the free 
electron mass. 
Driving such a material with $\lambda_0 = 10 \mu$m 
and leaving $F_0$ unchanged increases QSHHG by a factor of $5 \times 10^6$ over Fig. \ref{fig1}(b). 
Taking $N=8$ in Fig. \ref{fig1}(b) as an example, QSHHG converts $10^{11}$ squeezed vacuum photons into $10^2$ photons: a conversion efficiency of $10^{-9}$. Selecting the above material and laser wavelength therefore enables an increase of the conversion efficiency to $\approx 0.005$. Further optimizations of the pump laser parameters, such as an increased beam radius, 
and through the use of nanostructures to enhance the density of states \cite{sivis2017tailored, liu2018enhanced, vampa2017plasmon}, bring a conversion efficiency approaching unity within reach. 

The limit of near-unity conversion efficiency can be quite beneficial, as it would enable the frequency conversion of weak quantum optical states, such as Fock and entangled states 
(i.e. Bell states, etc...), to short wavelengths. Additionally, high conversion efficiency can be used to create potentially useful correlations and squeezing in the VUV to XUV wavelength regime. 
For example, if the SFG pathway for a single QSHH mode can be preferentially selected, Eq. (\ref{qshhq}) transitions into a 
two-mode sum-frequency operator. The resulting QSHH mode would be squeezed and highly efficient, as 
all photon pairs in the perturbation beam would get converted to that particular sideband \cite{vollmer2014quantum}. Sideband selection can potentially be obtained through resonances of a material
\cite{raz2012spectral, uzan2020attosecond} or a meta-surface, or through phase matching. 

Finally, the number of HH photons in solids and atomic gases is within the range observed in experiments 
\cite{lemieux2024photon}. Thus, our simple closed-form approach, despite its approximations, exhibits an
order of magnitude predictive power for conventional HHG. The accuracy can be further improved by adding 
absorption, more accurate models for the phase mismatch between laser and harmonics, by performing the
phase matching integrals exactly, and by accounting for the effect of the Coulomb potential in the 
Schr\"odinger equation.

\subsection{QSHH probability distribution}
\label{quasprob}

\noindent
In this section the single effective mode QSHH photon distribution probability is calculated. It is sufficient to use a two-mode wavefunction $\vert \varphi_m \rangle$ consisting of the quantum perturbation and a single harmonic sideband. Tracing out the other 
harmonic sidebands yields higher order corrections. 

The two-mode probability distribution is obtained from the wavefunction (\ref{nordeff}) as 
\begin{align}
P(m,n) = \left \vert \left \langle m, n \vert \varphi_m \right \rangle \right \vert^2 \mathrm{,}
\label{2modprob}
\end{align}
where $m,n$ refer to the photon number of the QSHH, and perturbative quantum mode, respectively. By summing over $n$, $P(m) = \sum_{n} P(m,n)$, and assuming an intense squeezed vacuum 
field $r \gg 1$, one obtains the QSHHG photon distribution, 
\begin{align}
P(m) \approx \frac{(2m-1)!!}{(2m)!!} \frac{\left( 2 \langle \hat{n} \rangle_{N} \tanh^{2}(r) \right)^m}{(1+2 \langle \hat{n} \rangle_{N})^{m+1/2}} \mathrm{.}
\label{probh}
\end{align}
The distribution (\ref{probh}) is approximately normalized, $\sum_m P(m) = 1/\sqrt{1+2 \vert \zeta_N \vert^2} \approx 1$, as $\vert \zeta_N \vert^2 \ll 1$. The approximate expression and exact 
numerical result are found to be in excellent agreement \cite{supp}. 

\begin{figure}[h]
\centering
\includegraphics[width=8.5cm]{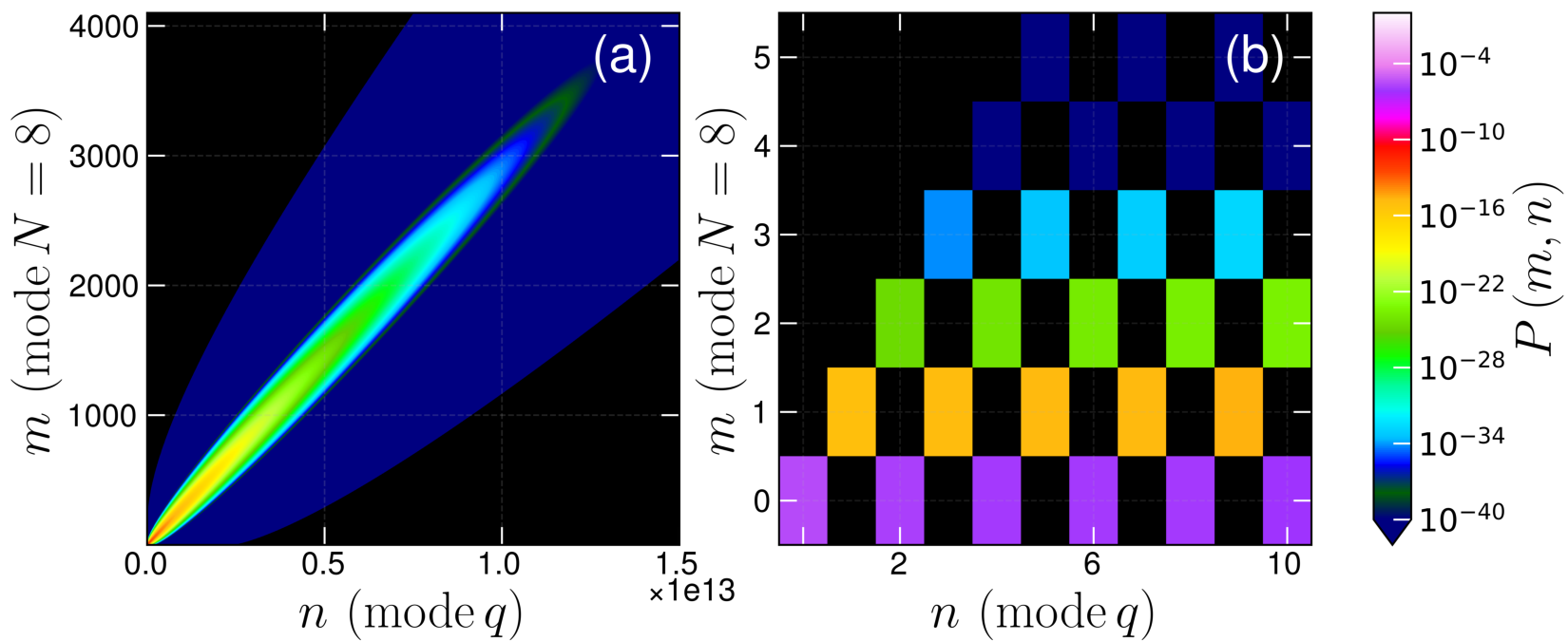}
\caption{(a) Two-mode probability of QSHHG versus photon number $n$ of the perturbative quantum mode and harmonic photon number $m$ for $N=8$; all parameters and $\langle \hat{n} \rangle_{N}(\theta=0)$
are from Fig. \ref{fig1}(c). (b) Close-up for photon numbers close to zero.}
\label{fig2}
\end{figure}

The two-mode distribution (\ref{2modprob}) is plotted in Fig. \ref{fig2}(a) for harmonic order $N=8$, see Fig. \ref{fig1}(c) for $\langle \hat{n} \rangle_{N}(\theta=0)$; (b) shows a close-up for photon 
numbers close to zero. The distribution has been calculated numerically from Eqs. (\ref{2modprob}) and (\ref{nordeff}). Clearly, harmonic and quantum modes are entangled, as only even-even or odd-odd
states are populated. Due to the large value of $r$, the probability extends to very high photon numbers in the quantum mode. The states $m>n$ are approximately zero, accurate to first order in $\vert 
\zeta_N \vert^2$. 

\begin{figure}[h]
\centering
\hspace*{-0.4cm}
\includegraphics[width=9.2cm]{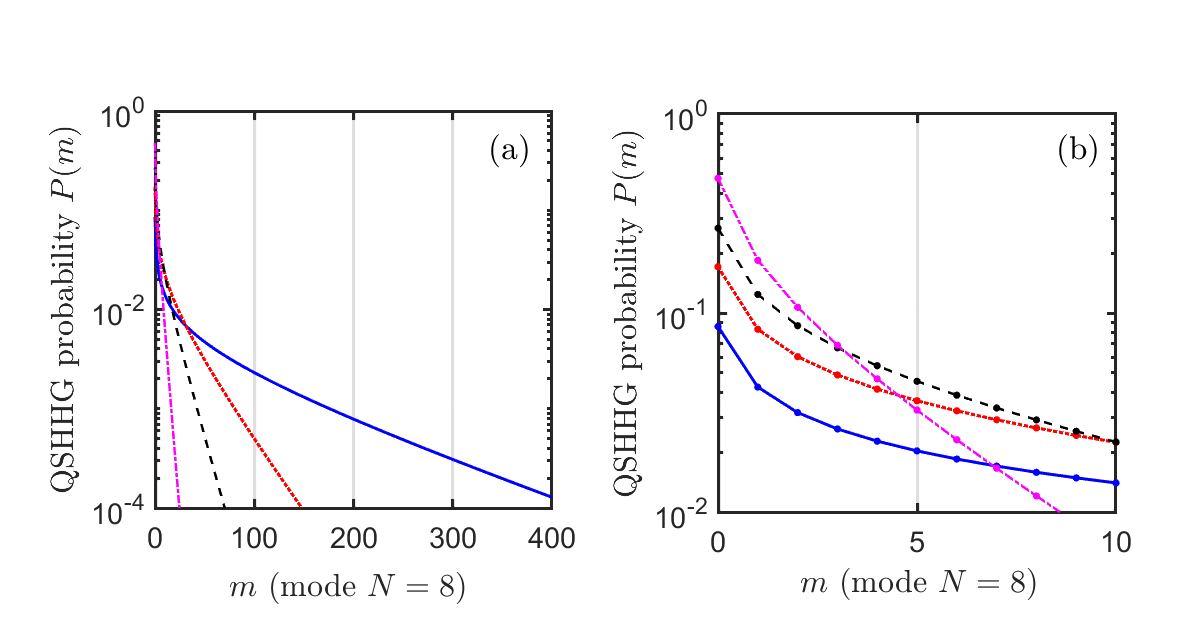}
\caption{(a) Probability of QSHHG versus harmonic photon number $m$ for $N=8$ (full blue line), $N=10$ (red dotted line), $N=12$ (black dashed line), $N=14$ (magenta dash-dotted line); $m$ refers to 
number of photons in the QSHH mode $N$; all parameters and $\langle \hat{n} \rangle_{N}(\theta=0)$ are from Fig. \ref{fig1}(c). (b) close-up for photon numbers close to zero.}
\label{fig3}
\end{figure}

In Fig. \ref{fig3}(a) the QSHHG probability $P(m)$ is plotted with a closeup in (b). The probability
(\ref{probh}) agrees with the probability of a squeezed vacuum state, with the distinction that every 
photon state is populated and not only the even states. This is a consequence of the two-mode even-even, 
odd-odd probability distribution. When traced over one of the modes, all number states of the remaining 
mode are populated and the non-classical features are washed out. Following the above approach, 
$g^{(2)}(0)$ and the quadrature are obtained, which indicate a super-Poissonian distribution and photon bunching. However no non-classical features, such as squeezing, sub-Poissonian statistics, or a negative Wigner distribution \cite{supp, lemieux2024photon} are indicated. In other words, for low conversion efficiency, all combinations of number states are populated and the non-classical features are traced out. However, in the case of unity conversion efficiency, where every BSV photon creates a QSHH photon, and for a single QSHH mode, only even QSHH number states are populated, thereby preserving the non-classical features \cite{vollmer2014quantum}. 

For more than one QSHH mode, the entanglement between modes gets more complicated. For example, with three modes, Taylor expansion of Eq. (\ref{nordeff}) yields operator terms $\sum_{mnj} (\hat{a}_{N}^{\dagger})^{2n+\eta_{n}} 
(\hat{a}_{N^{\prime}}^{\dagger})^{2m+\eta_{m}} (\hat{a}_{q}^{\dagger})^{j}$, where $\eta_{n}, \eta_{m} = 0,1$ for even, odd harmonic photon number states, respectively. The expansion coefficient of 
the squeezed vacuum operator is denoted by $j$ and the photon number of the quantum mode is given by $l = 2j+2n+2m+\eta_{n}+\eta_{m}$. As a result, $l$ can only be odd when only one of the harmonic modes
is odd: $\eta_{n} = 0,1$ and $\eta_{m} = 1,0$. When both harmonic modes are odd or even, then $l$ is even. While $l$ can 
be any positive integer, $n$ and $m$ consist of only even or odd integers for a given $l$,
thus resulting in a higher-dimensional checker board pattern similar to Fig. \ref{fig2}. Exploring multi-mode correlation presents an interesting avenue for future work. 
When the quantum mode is traced over in this three-mode case, all (even-even, odd-odd, and even-odd) states of the two harmonic modes will be populated. It is expected that this will also suppress non-classical properties and entanglement between harmonic modes, for low conversion efficiency.

The focus of the rest of the paper is on single-mode properties of QSHHG. In the following, we 
demonstrate how non-classical light can be obtained from QSHHG by projective measurements. This relies 
on the entanglement between quantum and harmonic modes. Projective measurements of entangled 
wavefunctions generally result in non-classical behavior \cite{gerry2023introductory}. 

\subsection{Projective measurement on perturbative quantum mode q}
\label{projq}

\noindent
Multi-mode entanglement requires measuring all sidebands simultaneously, or limiting emission to select 
sidebands only; otherwise, when entangled sidebands are not measured and traced out, mixed states are
generated and quantum properties destroyed \cite{gerry2023introductory}. Here we discuss the case of one 
and two sidebands. Our analysis starts with a single sideband, so that the wavefunction is limited to two 
modes. The number state of the perturbative quantum mode is determined as $\vert l \rangle_q$ by a
projective measurement. Determination of the resulting wavefunction starts with a Taylor expansion of the 
effective two-mode wavefunction (\ref{nordeff}) followed by a projection on 
$\vert l \rangle_q$. The projected wavefunction, $\left \vert \varphi_{N} \right \rangle =
\mathbin{_{q} \langle_{}} l \vert \varphi_m \rangle$, is found to be \cite{supp}, 
\begin{align}
& \left \vert \varphi_{N} \right \rangle \!=\! N_{\eta} (-e^{i \theta})^{l/2} \! \sum_{m=0}^{l_{\eta}/2} (-1)^m 
\frac{\sqrt{\alpha_{N}}^{2m+\eta}}{\sqrt{(2m+\eta)!}} \left \vert 2m+\eta \right \rangle_{N} \nonumber \\
& \alpha_{N} = \frac{l_{\eta} \vert \zeta_{N} \vert^2}{2 \beta_{N}}, \,\,\, l_{\eta} = l - \eta, \,\,\,\, \eta = 0,1 \, \mathrm{for \, even, odd} \, l \nonumber \\
& N_{ \eta = 0} \approx \frac{1}{\sqrt{\cosh(\vert \alpha_{N} \vert)}} \,\,\,\,\,\, N_{\eta = 1} \approx \frac{1}{\sqrt{\sinh(\vert \alpha_{N} \vert)}} \mathrm{.}
\label{qfix}
\end{align}
As seen in Fig. \ref{fig2}, the product $\hat{S}_m \hat{S}_q$ produces either even-even or odd-odd number states. As such, when the projected quantum number $l$ is even ($\eta=0$) or odd ($\eta=1$), 
the resulting harmonic wavefunction contains only even or odd states, respectively. 

From Eq. (\ref{qfix}), second order coherence and quadrature of the QSHH modes are calculated. They are a function of the projected quantum mode photon number $l_{\eta}$. 
Definitions and details on the calculation are given in the supplement; we find \cite{supp}
\begin{align}
& g^{(2)}_{N} = \frac{1}{\tanh^2(\vert \alpha_{\mkern-2mu N} \vert)} \,\,\,\,\,\,\,\, \mathrm{for} \, \eta = 0 \nonumber \\
& g^{(2)}_{N} = \tanh^2(\vert \alpha_{\mkern-2mu N} \vert) \,\,\,\,\,\,\,\,\, \mathrm{for} \, \eta = 1 \mathrm{}
\label{n2fixg2} 
\end{align}
with the second order coherence $g^{(2)} = (\langle \hat{n}^2 \rangle - \langle \hat{n} \rangle)/ \langle \hat{n} \rangle^2$, and 
\begin{align}
& \Delta X_{\,j N}^2 \!=\! \frac{1}{4} \left(1 \!+\! 2 \left(\vert \alpha_{\mkern-2mu N} \vert \tanh(\vert \alpha_{\mkern-2mu N} \vert) \!+\! (-1)^j \mathrm{Re}[\alpha_{\mkern-2mu N}] \right) 
\! \right) \! \mathrm{,} \,\, \eta = 0
\nonumber \\
& \Delta X_{\,j N}^2 \!=\! \frac{1}{4} \! \left(\! 1 \!+\! 2 \left( \frac{\vert \alpha_{\mkern-2mu N} \vert}{\tanh(\vert \alpha_{\mkern-2mu N} \vert)} 
\!+\! (-1)^j \mathrm{Re}[\alpha_{\mkern-2mu N}] \right) \! \right) \! \mathrm{,} \,\,\,\,\,\,\,\, \eta = 1
\mathrm{,} \label{n2fixquad} 
\end{align} 
where $j = 1,2$, the quadratures are $\hat{X}_j = 1/(2 i^{j-1}) (\hat{a} - (-1)^j \hat{a}^{\dagger})$, and the variances $ \Delta X_j^2 = \langle \hat{X}_j^2 \rangle - \langle \hat{X}_j \rangle^2$. All of 
the above equations were derived for $r \gg 1$, and  $l_{\eta} \gg 1$, and give excellent agreement with exact, numerical results in this limit \cite{supp}. Even for small $l_{\eta}$, the agreement 
is surprisingly decent.  

\begin{figure}[h]
\centering
\hspace*{-1.5cm}
\includegraphics[width=14cm]{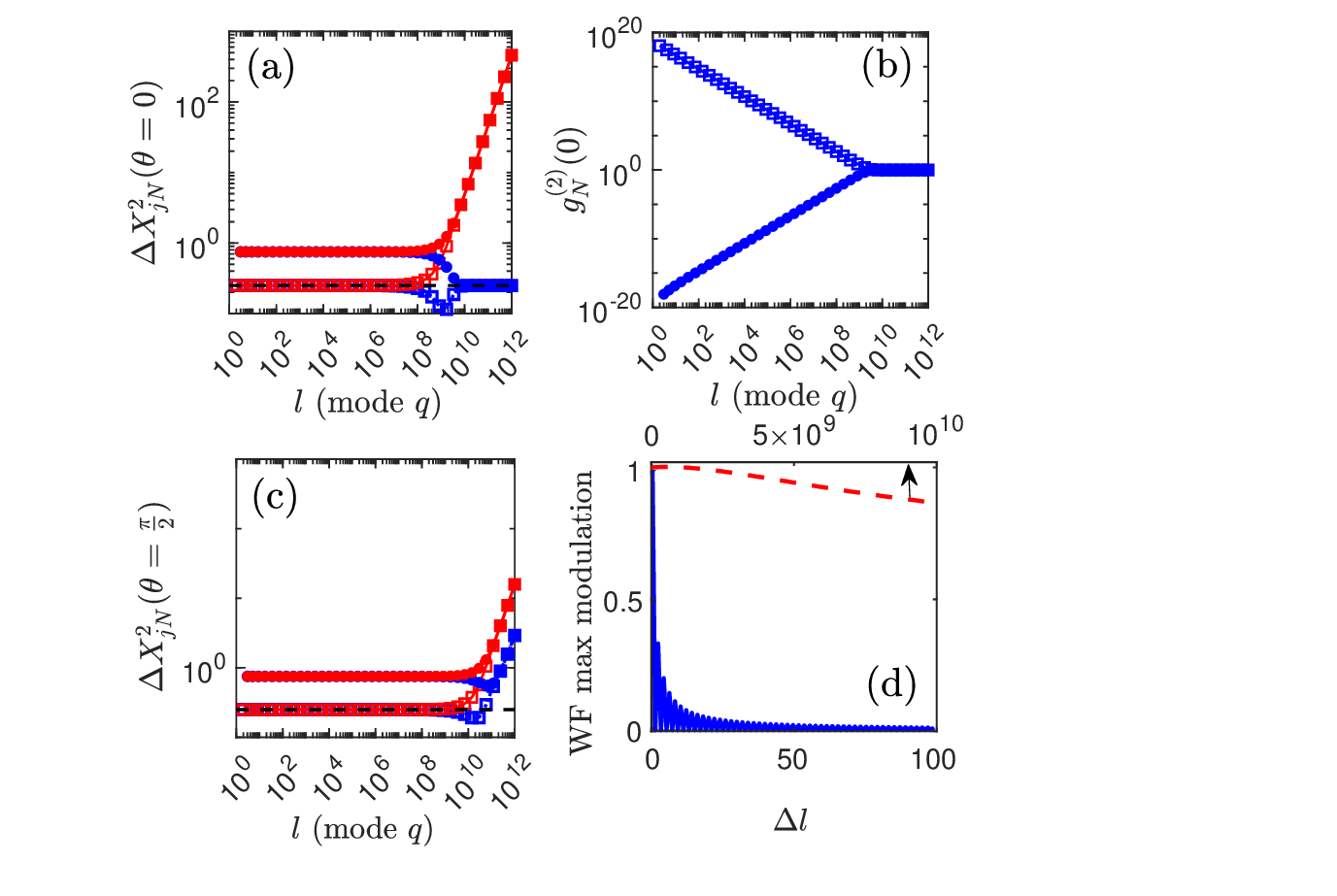}
\caption{Parameters are the same as in Fig.\ref{fig1}; (a-d) $\vert \zeta_N \vert^2(\theta=0)$ taken from $\langle \hat{n} \rangle_{N}$ for $N=8$ in \ref{fig1}(c), respectively. 
(a),(c) $\Delta X_{jN}^2(\theta=0,\pi/2)$ versus quantum photon number $l$, respectively; blue dotted, red full lines show $\Delta X_{jN}^2$ for $j=1,2$  respectively; symbols: in (a-c) squares and 
dots represent $\eta = 0,1$ (even, odd states), respectively; symbol spacing follows a $\log$ distribution; black dashed lines indicate quadrature variance of vacuum. (b) second order correlation
function $g^{(2)}(0,\theta=0)$ versus $l$. (d) Maximum modulation of Wigner function versus $l + \Delta l$, with $l = 10^{10}$ and $\Delta l$ the number state resolution; curves are normalized to the 
modulation at $\Delta l = 0$; blue solid line is for $\Delta l$ even and odd numbers, whereas red dashed line is for $\Delta l$ only even numbers.}
\label{fig4}
\end{figure}

The quadrature variances and the second order correlation function are plotted for $N=8$ as a function of quantum mode photon number $l$ in Figs. \ref{fig4}(a),(c) and (b), respectively. The parameter 
$\vert \alpha_N(\theta) \vert$ varies with $\theta$ for a given $l$. As such, quadrature variance and second order coherence are $\theta$-dependent. Panels (a,b) are for $\theta = 0$ and (c) is for 
$\theta = \pi/2$. The second order correlation function for $\theta = \pi/2$ is not shown, as it is only quantitatively slightly different. 

Note that the sum in the wavefunction (\ref{qfix}) has a finite limit $l_{\eta}$. As a result, it goes over into a Schr\"odinger cat state \cite{gerry2023introductory} in the limit $l_{\eta} 
\rightarrow \infty$, i.e. coherent states with only odd or even number states populated. In the opposite limit $l_{\eta} \rightarrow 0$ it behaves like a heralded number state which through 
projection is collapsed into a single photon state of the QSHH mode.

The states with only even ($\eta = 0$ squares) or only odd ($\eta = 1$ dots) photon numbers $l$ behave fundamentally different. In addition, in Figs. \ref{fig4}(a) and (c), blue and red represents the variances $j=1,2$. In the limit 
of even $l_{\eta=0} \rightarrow 0$, the QSHH
distribution has the characteristics of a vacuum number state which has a quadrature variance of $1/4$, see blue and red open squares in Fig. \ref{fig4}(a), and $g^{(2)}(0) \rightarrow \infty$, see open squares in \ref{fig4}(b). The odd states are dominated by a one photon number state which has a quadrature variance of $3/4$, see blue and red dots in Fig. \ref{fig4}(a) and $g^{(2)}(0) \rightarrow 0$, see dots in Fig. \ref{fig4}(b). As such, the odd states, primarily composed of a one photon number state, follow a sub-Poissonian statistics and are highly non-classical. In the intermediate regime, $l \sim 10^{9}-10^{10}$ for (a), moderate squeezing does occur for the even numbered states. In the high $l_{\eta}$ limit, even and odd states merge. All 
states have the $g^{(2)}(0) = 1$ of a coherent state which is typical for Schr\"odinger cat states. For $\theta=0$, the quadrature $\Delta X_1^2 \rightarrow 1/4$, whereas $\Delta X_2^2$ grows as a 
power function. For $\theta=\pi/2$, both quadrature variances grow as power functions. 

For the three-mode case, projecting on quantum mode $q$, will separate even-even and odd-odd number state populations for even $l$ from even-odd two-mode states for odd $l$. This results in an entangled two-harmonic mode state which will be subject to future research. 
However, measuring only one of the two sidebands traces out the other and this likely results in a mixed state, i.e. no quantum features, since the sidebands are entangled \cite{gerry2023introductory}. Therefore, all emitted sidebands ought to be measured.

The fact that the projected odd number state ranges from a single photon state to a Schr\"odinger cat state in dependence on $l_{\eta}$ opens the possibility of quantum engineering QSHHG. Experimental 
realization is complicated by two facts. (i) The probability of the single photon state is low. (ii) The wavefunction contains even and odd number states; due to the wide range of $l_{\eta}$, it is
difficult to resolve number states into single photons. As a result, even and odd states will mix and average out squeezing and the $g^{(2)}_{N}(0)$; for example, amplitude squeezing of the single photon state for low $l$ will be
lost. 

(i) We find from Fig. \ref{fig2}(b) that the probability of the small $n = l_{\eta}$ odd photon states is dominated by $m=1$ with a probability of $\approx 10^{-15}$. The probability scales as 
$\propto (\vert \zeta_{N} \vert^2)^{2m + \eta}/\cosh(r)$, and therefore drops fast with increasing $2m + \eta$. As the two-mode probability in Fig. \ref{fig2}(b) is approximately triangular, the one photon number state is essentially pure. Few photon number states $m$ offer less of an advantage; they
consist of a superposition of even or odd number states from $n=0$ or $n=1$ up to $n=m$, respectively. Their purity is comparable to that of a finite Schr\"odinger cat state.

We recall from our discussion of increasing the conversion efficiency of QSHHG below Eq. (\ref{rel_diff}) that it can be increased by more than six orders of magnitude by optimizing material and laser
parameters. We decrease $r$ so that the average number of photons, $\hat{n}_{N} = \vert \zeta_{N} \vert^2 \cosh^2(r)$ remains constant. As a result $\cosh(r)$ decreases by three orders of magnitude, and 
the probability of creation of a one photon state increases by nine orders of magnitude to $10^{-6}$. If in addition other parameters are optimized, by for example using a wider pump beam radius, adding
another 2 orders of magnitude, a probability for generating a single photon state of about $10^{-4}$ appears achievable. For a MHz laser system, this would mean the generation of a single photon state
every $10$ms. 

(ii) Techniques for resolving the number of photons are limited to $l \sim 10$ \cite{schmidt2018photon}. This gives experimental access to single-photon and few odd-photon states and presents a path to shift
the generation of pure heralded single-photon states towards the XUV. In the opposite limit of Schr\"odinger cat states, $ l \gg 1$, projecting on the number state with large $l$ presents a considerable problem for existing technologies. Schr\"odinger cat states exhibit characteristic interferences near the zeros of the quadratures, with opposite phase shifts for odd- and even-parity states and with varying fringe spacing for different average number of photons. With increasing uncertainty in the photon number resolution, these fringes will be washed out, see Fig. \ref{fig4}(d), where $l=10^{10}$, $\Delta l$ represents the photon number resolution, and the Wigner function is averaged over $\Delta l$. Snapshots of the Wigner function for various $\Delta l$ are shown in the supplement \cite{supp}. 
We show two cases, one with $\Delta l$ running over even and odd numbers (blue full line), and one assuming that parity measurement is possible with $\Delta l$ only extending over even numbers. Without parity measurement, the fringes average out over $100$ number states, for even $\Delta l$ and they 
vanish immediately for odd $\Delta l$, as even ($l$) and the next odd state ($l+1$) are phase shifted 
by $\pi$. On the other hand, with the parity measurement, averaging over $10^{10}$ (even only) number states is possible with small losses in the fringe visibility. However, preserving the parity of the quantum mode for such large $l$ is practically impossible, as loosing one photon in $\sim10^{10}$ would suffice to toggle the parity of the state. The conversion efficiency would have to be increased to levels higher than the typical optical losses of an experiment ($\sim 1\%$) for the parity to be maintained and for the cat state to be measurable. Even without knowledge of parity, a measurement of $g^{(2)}(0) \rightarrow 1$ and of the quadratures' excess 
noise by projecting over a range of photon numbers would provide an indication of the generation of the cat states. The quadratures's noise can be measured in the (few photons) quantum sidebands with
 homodyne interferometry \cite{lvovsky2015squeezed} or, potentially, by extending \textit{in-situ} techniques to the single shot \cite{dudovich2006measuring}.


\subsection{Projective measurement on QSHH mode $N$}
\label{projN}

\noindent
The analysis is performed for Eq. (\ref{nordeff}) with one QSHH mode, i.e. under the assumption that only one sideband is generated. Tracing out the other sidebands in a multi-mode scenario will modify the parameters $\zeta_N$ and $\beta_N$, but will leave the two-mode wavefunction unchanged otherwise. In a projective measurement, the number of photons $m$ in the effective QSHH mode $N$ is measured. The resulting wavefunction for 
the quantum mode $q$ depends on $m$ and is calculated by projecting an effective number state on the wavefunction (\ref{nordeff}) limited to two modes, $\vert \varphi_q \rangle = \mathbin{_{N} \langle_{}} m \vert \varphi_m \rangle $. 
This yields 
\begin{align}
\vert \varphi_q \rangle = N_m \sum_{l=0}^{\infty} (-1)^l \frac{\sqrt{(2l+m)!}}{l!} \beta_{N}^l \left \vert 2l+m \right \rangle_{q} \mathrm{.}
\label{kfix} 
\end{align}
The norm is given by 
\begin{align}
\frac{1}{N_m^2} &\! \approx \frac{\cosh(r)}{\sqrt{1 + 2 \langle \hat{n} \rangle_{\!N}} } \Biggl[ \! (2m-1)!! \! \left( \frac{\sinh(r)}{1 \!+\! 
2 \langle \hat{n} \rangle_{\!N}} \! \right)^m 
\nonumber \\ 
& + \frac{l}{2} (3m-1) (2m-3)!! \left( \frac{\sinh(r)}{1 + 2 \langle \hat{n} \rangle_{\!N}} \right)^{m-1} \Biggr] \mathrm{.}
\label{kfixnrm} 
\end{align}

With the above wavefunction second order coherence and quadrature variances are calculated; for definitions and details see the supplement \cite{supp}. We obtain for the second order coherence 
\begin{align}
g^{(2)}_{N}(0) \approx 1 + \frac{2}{2m+1} \mathrm{.}
\label{kfixg2} 
\end{align}
The quadratures ($j=1,2$) are found to be 
\begin{align}
\Delta X_{\!jN}^2 & \approx \frac{m(m-1)}{2(2m-1)} \!\! \left(1 + (-1)^j  \cos(\theta) \frac{1 + \vert \zeta_N \vert^2}{\tanh r} \right) \nonumber \\
& + \frac{(2m\!+\!1)}{4} \frac{A_j(r,\theta) + \vert \zeta_N \vert^2 (1 \!+\! \langle \hat{n} \rangle_{\!N} \!)}{1+ 2 \langle \hat{n} \rangle_{\!N}} \mathrm{} 
\label{kfixqu}
\end{align}
with $A_j(r, \theta) = \cosh^2 \!r \!+\! \sinh^2 \!r \!+ \! 2 (-1)^j \! \cosh r \sinh r \cos(\theta)$ and $\langle \hat{n} \rangle_{N} = \cosh^2 \!r \vert \zeta_N \vert^2$. 
For the case $\theta = 0$, $\tanh r \approx 1$, $\langle \hat{n} \rangle_{N} \gg 1$ one obtains
\begin{align}
\Delta X_{\!jN}^2 \! & \approx \! \frac{2m+1}{4} \frac{ (\cosh(r) \!+\! (-1)^j \! \sinh(r))^2}{1 \!+\! 2 \langle \hat{n} \rangle_{\!N}} \nonumber \\
& + \frac{\vert \zeta_N \vert^2}{4} \!\! \left(\! 1 \!+\! \frac{1}{2(2m\!-\!1)} \! \right) \! \mathrm{.} 
\label{kfixqux} 
\end{align}
We see that the quadrature consists of two contributions; the first term contains the quadrature of a squeezed vacuum state, but has a prefactor that depends on $m$ and 
$\langle \hat{n} \rangle_{\!N}$. The second term is commonly not part of BSV states. Both modifications come from the mode mixing term $\hat{a}_{N}^{\dagger} \hat{a}_{q}^{\dagger}$ in Eq. 
(\ref{nordeff}). Finally, all of the above equations give excellent agreement with the exact, numerical results 
\cite{supp}.

\begin{figure}[h]
\centering
\hspace*{-1cm}
\includegraphics[width=10.4cm]{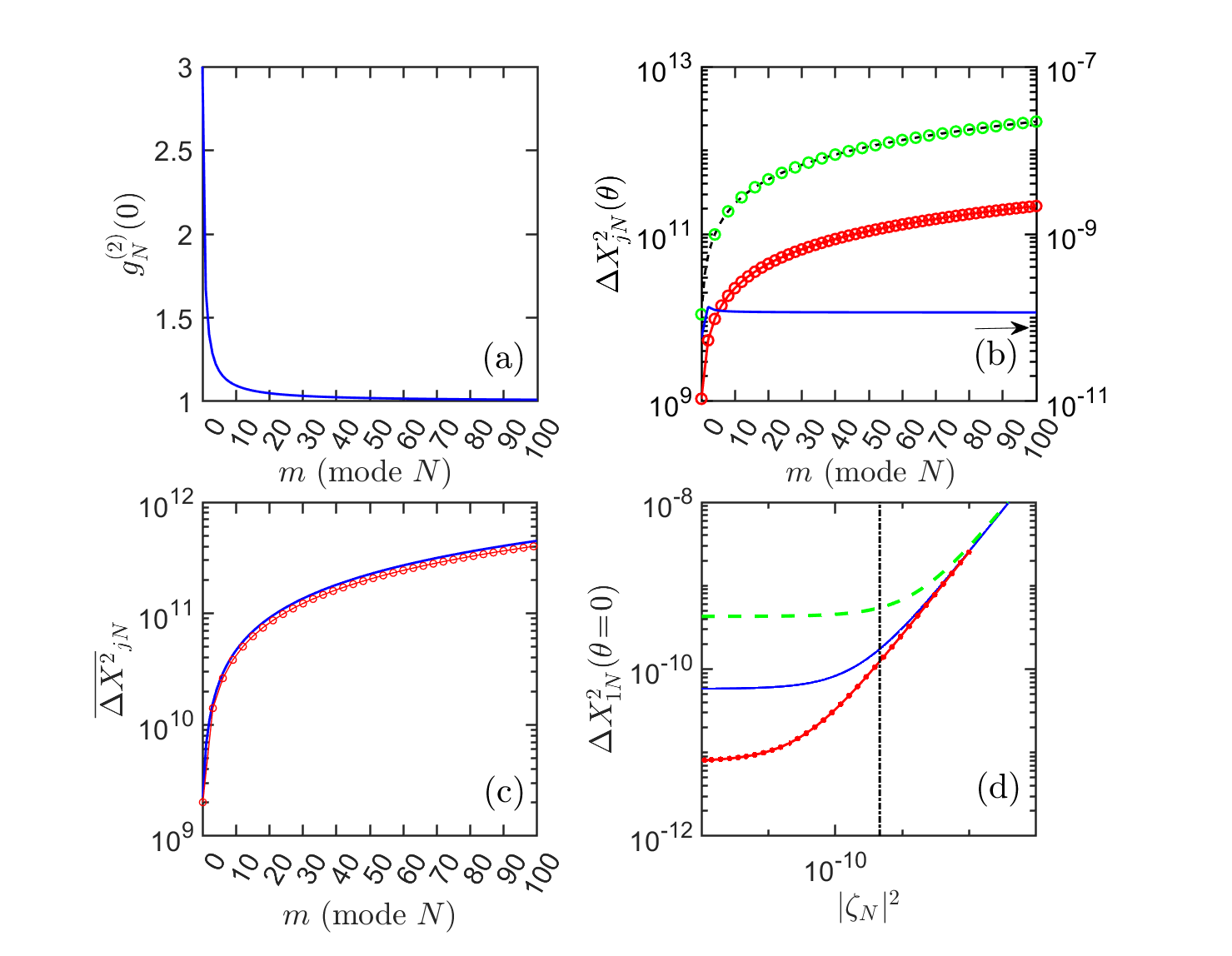}
\vspace*{-0.1cm}
\caption{(a) second order correlation $g^{(2)}_{N}(0)$ versus $m$; independent of all other parameters. (b) Quadrature variances $\Delta X_{\,jN}^2(\theta)$ ($j=1,2$) for squeezed vacuum angle 
$\theta$ versus harmonic photon number $m$; evaluated for the effective QSHH order $N=8$; parameters from Fig. \ref{fig1}(b); blue line, red circles for $\theta=0$, $j=1,2$, respectively; black dashed line, green circles (on top) for $\theta=\pi/2$, $j=1,2$, respectively. (c) same as (b), but angle ($\theta$) averaged quadrature variances for $N=8$; blue line, red line with open circles for $j=1,2$, respectively. (d) $\Delta X_{\,1}^2(\theta=0)$ evaluated from Eq. (\ref{kfixqux}) versus $\vert \zeta_N \vert^2$ for $m=50$ and for $r=12.4$ (green dashed), $r=13.4$ 
(full blue), $r=14.4$ (red line with dots); dash-dotted line indicates $\vert \overline{\zeta}_N \vert^2$ of (b) for $N=8$. } 
\label{fig5}
\end{figure}

The wavefunction in Eq. (\ref{kfix}) represents an $m$-photon added squeezed vacuum state \cite{kun2010nonclassicality}. For $m = 0$, $g^{(2)}_{N}(0) =3$ takes the value of a squeezed vacuum state, see Fig. \ref{fig5}(a). Further, the product of quadrature variances in Fig. \ref{fig5}(b) for $m=0$ and $\theta=0$ gives $1/16$, the minimum uncertainty. However, the maximum degree of squeezing, occurring for $m=0$ is reduced from that of the input state, which is $\Delta X_1^2=\frac{1}{4}e^{-2r}\simeq 5 \times 10^{-13}$. This can be understood from the fact that mixing between harmonic and quantum modes in QSHHG changes $\beta \rightarrow \beta_N$, so that in Eq. (\ref{kfixqux}) a residual term $\propto \vert \zeta_N \vert^2$ remains, even for $m=0$. This term is positive and reduces squeezing, although minimum uncertainty is sustained. Finally, due to mode mixing, the quadrature variances for $m \ne 0$ are different from a regular squeezed state, see blue line and red dots in Fig. \ref{fig5}(b) for $\Delta X_{\,jN}^2(\theta=0)$ with $j=1,2$, respectively. Both quadratures increase with $m$, making the state larger than a minimum uncertainty state.

For $\theta=\pi/2$ squeezing disappears, like for a squeezed vacuum beam, see the black dashed line and green circles in Fig. \ref{fig5}(b) for $j=1,2$, respectively. The angle averaged quadrature 
variance also shows no squeezing, in agreement with a regular squeezed vacuum beam; however 
$\overline{\Delta X_{\,jN}^2}$ with $j=1,2$ need not necessarily be the same, as $\vert \zeta_N(\theta) \vert^2$ is 
not the same for $\theta$ and $\theta + \pi$, see Fig. \ref{fig1}(c). 
Finally, Eq. (\ref{kfixqux}) is evaluated for a range of $\vert \zeta_N \vert^2$ for three values of $r$ and for $m=50$; the quadrature variance $\Delta X_{\,1N}^2(\theta=0)$ is 
plotted versus $\vert \zeta_N \vert^2$ in Fig. \ref{fig5}(d). The black dash-dotted line indicates the value of $\vert \zeta_N \vert^2$ in \ref{fig5}(b). The squeezed vacuum parameter exerts large
influence for small $\vert \zeta_N \vert^2$ for which the first term in Eq. (\ref{kfixqux}) is dominant. This influence disappears for larger $\vert \zeta_N \vert^2$, where the second term in 
Eq. (\ref{kfixqux}) dominates. The linear growth highlights that high conversion efficiency compromises the potential for squeezing. 

Photon-added squeezed vacuum states have demonstrated an ability to measure phase closer to the Heisenberg limit than other quantum states, such as squeezed vacuum states \cite{guo2018improving}. Thus,
measuring such states during QSHHG seems important. As the projection is done on the (weak) QSHH mode, photon number resolution is less of an issue than in the previous section. The photon added 
squeezed vacuum state remains a macroscopic state whose quadratures can be measured with conventional homodyne detection \cite{lvovsky2015squeezed} or an extension of attosecond techniques such as TIPTOE \cite{park2018direct} and nonlinear photoconductive field sampling \cite{sederberg2020attosecond} of quantum fields. Examples of Wigner functions for this state are shown in Fig. (S5) of 
\cite{supp}. The only difficulty is resolving the squeezed quadrature, which can be technically challenging due to saturation of
balanced detectors at the high photon fluxes in the quantum mode. However, as discussed above the efficiency of QSHHG can be
substantially increased, which allows the use of quantum beams with a much smaller amount of photons. Aside from these technical
challenges, an indication of photon added squeezed vacuum states could be obtained experimentally, by measuring the increase of the 
variance of the quadratures and decrease of $g^{(2)}_{N}(0)$ with increasing $m$. 

\section{Conclusion}

\noindent
High-harmonic generation is a process that can emit coherent XUV radiation, when driven by sufficiently intense lasers. For classical driving lasers, commonly classical XUV radiation is emitted. When the classical pump laser is replaced by a quantum field, such as a bright squeezed vacuum, properties of high-harmonic generation, such as cutoff and photon statistics (bunching), are modified, however non-classical properties have not been demonstrated yet. In addition, scaling quantum HHG into the XUV regime appears to be challenging. 

Quantum sideband high harmonic generation serves as an alternative approach that leverages the strength of readily available intense classical fields to generate the short-wavelength high harmonics, and the quantum properties of a weak perturbing beam to engineer the quantum-optical properties of the harmonics. This imposes less stringent restrictions on the brightness of the quantum perturbation and as such, widens the range of suitable quantum sources. 

In this work, a comprehensive theoretical framework of quantum sideband high-harmonic generation has been introduced. It serves as a 
foundation for designing and developing quantum sideband high-harmonic generation as a short-wavelength attosecond source for quantum light 
generation. Our theoretical analysis has revealed that quantum sideband high harmonic generation creates entanglement between the 
harmonic sideband modes and between the harmonic sidebands and the quantum perturbation. The entanglement allows to quantum engineer high
harmonic generation via projective measurements. As a result, a variety of states commonly used in quantum information theory can be generated,
such as heralded number states, Schr\"odinger cat states, and photon added squeezed vacuum states. These states will find applications in
extending quantum sensing, quantum imaging, and quantum information science to the XUV regime. 

Challenges to realize theses states experimentally and potential pathways to overcome them have been discussed. The biggest challenges are (i) the many-mode entanglement between all the sidebands and the BSV and (ii) the low conversion efficiency, which requires BSV beams with a large number of photons; however, still significantly less than what needed to drive high harmonic emission with the BVS field alone. 

(i) Multi-mode entanglement requires measuring all sidebands simultaneously to avoid generating a mixed state, when projecting on the BSV. Strategies to monochromatize the emission spectrum were discussed, such as through the use of resonances or phase matching. On the other hand, multi-mode entanglement provides an avenue for creating multi-mode correlations, an interesting avenue for future work. 

(ii) The great disparity in photon numbers between sidebands and quantum mode limits the ability to generate, preserve and measure the quantum-optical states, whether generated by direct upconversion or by a projective measurement. This is a common issue when dealing with bright quantum-optical states, and an active area of research. 
Our simulations reveal that the conversion efficiency scales very favorably with the laser parameters (intensity, wavelength), and the effective mass of the electron, and that near-unity conversion efficiency can potentially be achieved with the help of resonant metasurfaces and optimized phase matching conditions. This boost in efficiency would unlock the potential of high-order sideband generation for the creation of the quantum-optical states predicted in this work.

\bibliographystyle{apsrev4-2} 
\bibliography{manuscript_GV} %

\clearpage

\end{document}


\title{Quantum engineering of high harmonic generation: supplementary material}

\author{N. Boroumand}
\affiliation{Department of Physics, University of Ottawa, Ottawa, Ontario K1N 6N5, Canada}
\author{A. Thorpe}
\affiliation{Department of Physics, University of Ottawa, Ottawa, Ontario K1N 6N5, Canada}
\author{G. Bart}
\affiliation{Department of Physics, University of Ottawa, Ottawa, Ontario K1N 6N5, Canada}
\author{L. Wang}
\affiliation{Department of Physics, University of Ottawa, Ottawa, Ontario K1N 6N5, Canada}
\author{D. N. Purschke}
\affiliation{Joint Attosecond Science Laboratory, National Research Council of Canada and University of Ottawa, Ottawa, Ontario, K1A 0R6, Canada}
\author{G. Vampa}
\email[]{gvampa@uottawa.ca}
\affiliation{Joint Attosecond Science Laboratory, National Research Council of Canada and University of Ottawa, Ottawa, Ontario, K1A 0R6, Canada}
\author{T. Brabec}
\email[]{brabec@uottawa.ca}
\affiliation{Department of Physics, University of Ottawa, Ottawa, Ontario K1N 6N5, Canada}

\maketitle


\section{Photon wavefunction}

\noindent
The semiclassical model of intense laser atom interaction treats high harmonic generation (HHG) and intense pump laser as a classical electric field \cite{lewenstein1994theory}. Here, the model is 
generalized to treat high harmonic (HH) modes and quantum sideband (QSHH) modes quantum optically. Quantum sideband high harmonic generation (QSHHG) represents the generation of HH sidebands through
mixing of a quantum field, such as bright squeezed vacuum (BSV), with HHG. The intense pump laser modes are treated classically, and back-action of high harmonic (HH) modes on laser modes 
is neglected. Our analysis starts from the Schr\"odinger equation in length gauge, 
\begin{align}
i \hbar \partial_{t} \Psi = \hat{H} \Psi = \left( \hat{H}_m  + \hat{H}_f + \hat{H}_{i} \right) \Psi \mathrm{,}
\label{schroed}
\end{align}
where $ \hat{H}_m = \hat{\mathbf{p}}^2/(2m) + V(\mathbf{r})$ is the matter Hamiltonian, $\mathbf{p} = i\hbar \boldsymbol{\nabla}_{\mathbf{r}}$ is the momentum operator, 
$ \hat{H}_{f} = \sum_{\kappa} \hbar \omega_{k} \left(\hat{a}^{\dagger}_{\kappa} \hat{a}_{\kappa} \right)$ is the field Hamiltonian, 
and $ \hat{H}_{i} = \sum_j \mathbf{r} |e| \hat{\mathbf{F}}(\mathbf{x}_j)$ is the interaction Hamiltonian. Electron charge and mass are denoted by $e,m$, and $\mathbf{r}, 
\mathbf{x}_j$ represent the relative and center of mass coordinates of atom $j$. The atom index is not important for the microscopic part of the derivation and will not be given explicitly. 
It will be used for macroscopic HHG and QSHHG taking into account propagation effects. Further, $\hat{a}^{\dagger}_{\kappa}, \hat{a}_{\kappa}$ are the creation and annihilation operators of 
a light mode with polarization $\mathbf{e}_{\kappa}$ and wavevector $\mathbf{k} = k \mathbf{n}$; $k = \omega_k/c$, and $\mathbf{n}$ is the unit vector along propagation direction. For brevity,  
we use the multi-index $\kappa \equiv \mathbf{k}s$, where $s=1,2$ is the polarization index. Finally, the electric field and vector potential operators in dipole approximation, 
\begin{subequations}
\label{ea}
\begin{align}
\hat{\mathbf{F}}(\mathbf{x}) & = i \sum_{\kappa} \eta_{k} \omega_{k} \mathbf{e}_{\kappa} \left( \hat{a}_{\kappa} e^{i \mathbf{k} \mathbf{x}} - 
\hat{a}^{\dagger}_{\kappa} e^{-i \mathbf{k} \mathbf{x}} \right) 
\label{vece} \\
\hat{\mathbf{A}}(\mathbf{x}) & = \sum_{\kappa} \eta_{k} \mathbf{e}_{\kappa} \left( \hat{a}_{\kappa} e^{i \mathbf{k} \mathbf{x}} + 
\hat{a}^{\dagger}_{\kappa} e^{-i \mathbf{k} \mathbf{x}} \right) \text{} \label{veca}
\end{align}
\end{subequations}
do not depend on $\mathbf{r}$. Here, $\eta_k^2 = \hbar/(2\omega_k \varepsilon_0 V)$, $\varepsilon_0$ is the vacuum permittivity, and $V$ the quantization volume. 

Equation (\ref{schroed}) is solved by inserting the Ansatz 
\begin{align}
& \vert \Psi \rangle = \vert 0 \rangle \otimes \vert \phi_0(t) \rangle + \int \!\! d^3\mkern-2mu p \, \vert \mathbf{p} \rangle \otimes 
\vert \phi_{\mathbf{p}}(t) \rangle \mathrm{,}
\label{ansatz} 
\end{align}
where $\vert 0(\mathbf{r}) \rangle$ is the matter ground state with binding energy $E_0$; continuum states are approximated by plane wave states $\vert \mathbf{p} \rangle = 
1/(2\pi \hbar)^{3/2} \exp((i/\hbar) \mathbf{p} \mathbf{r})$, and $\phi_l(t)$ ($l = 0,p$) are the corresponding ground and continuum state photon wavefunctions. The strong field approximation is 
used which is based on two assumptions. First, the effect of the Coulomb potential on the continuum states is neglected \cite{lewenstein1994theory}. Second, ground state and continuum states are 
assumed to be approximately orthogonal, $\langle \mathbf{p} \vert 0 \rangle \approx 0$. 

The matter wavefunctions are integrated out by projecting the functionals $\langle 0 \vert$, $\langle \mathbf{p} \vert$ on Eq. (\ref{schroed}). This results in two coupled equations 
for the photon wavefunctions $\phi_0(t), \phi_{\mathbf{p}}(t)$. These equations are transformed by using the Ansatz $\vert \phi_l \rangle = \hat{U}_{i} \hat{U}_{v} \hat{D}_{\alpha} \vert \varphi_l 
\rangle $ ($l = 0, \mathbf{p}$) \cite{gorlach2020quantum}. The first operator $\hat{D}_{\alpha} = \prod_{\kappa} \hat{D}(\alpha_{\kappa})$ removes the strong laser field from the photon wavefunctions. 
Here, $\kappa$ runs over the modes of the classical pump field, and 
\begin{align}
\hat{D}(\alpha_{\kappa}) = \exp(\alpha_{\kappa}(t) \hat{a}_{\kappa}^{\dagger} - \alpha_{\kappa}^*(t) \hat{a}_{\kappa})
\label{shift}
\end{align}
is the coherent state operator of mode $\kappa$ with $\alpha_{\kappa}(t) = \alpha_{\kappa} \exp(-i \omega_k t)$. The second operator $\hat{U}_{v} = \exp(i\vert e \vert/\hbar\mathbf{r} \hat{\mathbf{A}})$ 
transforms the Schr\"odinger equation from length to velocity gauge. Finally, the third (interaction representation) operator $\hat{U}_{i} = \exp(-i/\hbar \hat{H}_f t)$ eliminates the field 
Hamiltonian and makes the photon operators time dependent. Working out the transformed equations of motion yields 
\begin{subequations}
\label{eqtldphi}
\begin{align}
& i \hbar \partial_t \vert \varphi_0 \rangle = E_0 \vert \varphi_{0} \rangle + \int \!\! d^3\mkern-2mu p \, \mathbf{d}^*\!(\mathbf{p}_t) \, 
\vert e \vert \tilde{\mathbf{F}}(t) \, \vert \varphi_{\mathbf{p}} \rangle  
\label{eqtldphi0} \\
& i \hbar \partial_t \vert \varphi_{\mathbf{p}} \rangle \! = \! \left(\frac{\mathbf{p}_t^2}{2m} \!+\! \frac{|e|}{m} \mathbf{p}_t \hat{\mathbf{A}}(t) \right) 
\vert \varphi_{\mathbf{p}} \rangle \! + \! \mathbf{d}(\mathbf{p}_t) \, \vert e \vert \tilde{\mathbf{F}}(t) \, \vert \varphi_0 \rangle \mathrm{,} 
\label{eqtldphip}
\end{align}
\end{subequations}
where $\mathbf{p}_{\mkern-1.5mu t} = \mathbf{p} + |e|\boldsymbol{\mathcal{A}}(t)$. Further, $\mathbf{d}(\mathbf{p}) = \langle \mathbf{p} \vert \mathbf{x} \vert 0 \rangle = d_0 \mathbf{p} / 
(\mathbf{p}^2/(2m) + E_0)^3$  is the transition dipole moment between ground and continuum plane wave states
\cite{lewenstein1994theory} with $d_0$ specified in the manuscript. Note that $\hat{D}_{\alpha}$ and $\hat{U}_i$ transform the 
electric field and vector potential operators into $(\hat{D}_{\alpha}(t) \hat{U}_i)^{\dagger} \hat{\mathbf{F}} 
\hat{D}_{\alpha}(t) \hat{U}_i = \tilde{\mathbf{F}}(t) = \boldsymbol{\mathcal{F}}(t) + \hat{\mathbf{F}}(t)$ and 
$(\hat{D}_{\alpha}(t) \hat{U}_i)^{\dagger} \hat{\mathbf{A}} \hat{D}_{\alpha}(t) \hat{U}_i = \boldsymbol{\mathcal{A}}(t) + \hat{\mathbf{A}}(t)$; 
$\boldsymbol{\mathcal{F}}(t)$ and $\boldsymbol{\mathcal{A}}(t)$ are the classical intense laser electric field and vector potential with center frequency $\omega_0$ and wavevector $\mathbf{k}_0$. 
All fields are time dependent in the interaction picture. Further, as the classical fields are much stronger than the quantum fields, $\boldsymbol{\mathcal{A}}^2$ is retained in 
Eq. (\ref{eqtldphi}), while $\hat{\mathbf{A}}^2$ is neglected. Although Eq. (\ref{eqtldphi}) still contains the quantum operators of the classical pump field modes, they are disregarded in 
the following derivation. Finally, all fields depend on the position of atom $\mathbf{x}_j$ at which HHG is taking place; this is not explicitly stated. 

For integration of Eq. (\ref{eqtldphip}) the following relation is useful, $\partial_t \exp(\hat{B}(t)) = (\partial_t \hat{B} + (1/2) [\hat{B}(t), \partial_t \hat{B}] ) 
\exp(\hat{B}(t))$, which is valid as long as the commutator gives a c-number. Here, the commutator results in a small phase term which is neglected. We obtain
\begin{align}
& \vert \varphi_{\mathbf{p}}(t) \rangle = -i \int_{t_0}^{t} \!\!\! dt^{\prime} \, \hat{c}_{\mathbf{p}}(t^{\prime}) \, \bigl \vert \varphi_0(t^{\prime}) \bigr \rangle \nonumber \\
& \hat{c}_{\mathbf{p}}(t) = \tilde{\Omega}(t) \hat{D}_{\sigma}^{\dagger}(t) \exp \left( i S(t) \right) \mathrm{.}
\label{dtldphip} 
\end{align}
where $t_0$ is some initial time. Further, $\tilde{\Omega}(t) = 
\Omega(t) + \hat{\Omega}(t) = (\vert e \vert / \hbar) \mathbf{d}(\mathbf{p}_{t}) \tilde{\mathbf{F}}(t)$ is a generalized Rabi frequency that consists of a classical part $\Omega(t) = 
(\vert e \vert /\hbar) \mathbf{d}(\mathbf{p}_t) \boldsymbol{\mathcal{F}}(t) $, and an operator part $\hat{\Omega}(t) = \sum_{\kappa} \hat{\Omega}_{\kappa}(t)$ with 
\begin{align}
\hat{\Omega}_{\kappa}(t) = -i \Omega_{\kappa}(t) \left(  \hat{a}^{\dagger}_{\kappa} e^{i \omega_{k} \tau } \!-\! 
\hat{a}_{\kappa} e^{-i \omega_{k} \tau } \right) \mathrm{.}
\end{align}
Here, $ \Omega_{\kappa}(t) = (\vert e \vert /\hbar) \eta_k \omega_k (\mathbf{e}_{\kappa} \mathbf{d}(\mathbf{p}_t) )$, and $\tau = t - (\mathbf{n} \mathbf{x}_j) / c$. The classical action integral 
is given by 
\begin{align}
S(t) = \frac{1}{\hbar} \int_{t_0}^{t} \! \left( \frac{1}{2m} \mathbf{p}^2_{\mkern-1.5mu \tau} + E_0 \right) d\tau \mathrm{.}
\label{S}
\end{align}
Finally, the continuum electron is dressed with a displacement operator $\hat{D}^{\dagger}_{\sigma}(t) = \prod_{\kappa} \hat{D}^{\dagger}( \sigma_{\kappa}(t))$; for a definition of 
$\hat{D}$ see Eq. (\ref{shift}). Here, $\hat{D}^{\dagger} \bigl( \sigma_{\kappa}(t) \bigr) = \exp(-\hat{\sigma}_{\kappa})$ with $\hat{\sigma}_{\kappa} = \sigma_{\kappa}(t) \hat{a}^{\dagger}_{\kappa} 
- \sigma^*_{\kappa}(t) \hat{a}_{\kappa}$ and 
\begin{align}
& \sigma_{\kappa}(t) = \! \frac{\vert e \vert E_v}{\hbar} \overline{\sigma}_{\kappa}(t) e^{i \omega_{k} \tau } \mathrm{,} \nonumber \\ 
& \overline{\sigma}_{\kappa}(t) = \! 
- \frac{i}{\omega_k} \! \int_{t_0}^{t} dt^{\prime}  \left( \mathbf{e}_{\kappa} \mathbf{v}_{t^{\prime}} \right)  
e^{-i \omega_{k} (t - t^{\prime})} \mathrm{.}
\label{sig}
\end{align}
For $\omega_k \gg \omega_0$, $\overline{\sigma}_{\kappa} \approx (\mathbf{e}_{\kappa} \mathbf{v}_{t}) / \omega_k^2$ where $\mathbf{v} = \mathbf{p}/m$ is the electron velocity. This assumes 
that the contribution of the integral at time $t_0$ can be neglected. 

Inserting Eq. (\ref{dtldphip}) into Eq. (\ref{eqtldphi0}) results in an integro-differential equation for $\vert \varphi_0 \rangle$, 
\begin{align}
\partial_t \vert \varphi_0(t) \rangle = - \int \!\! d^3\mkern-2mu p \, \hat{c}_{\mathbf{p}}^{\dagger}(t) \!\! \int_{t_0}^{t} \!\!\! dt^{\prime} \, 
\hat{c}_{\mathbf{p}}(t^{\prime}) \, \vert \varphi_0(t^{\prime}) \rangle \mathrm{.}
\label{dtldphi0} 
\end{align}
In the weak depletion limit, $\vert \varphi_0 \rangle$ can be pulled out of the integral on the right hand side of Eq. (\ref{dtldphi0}) by integration by parts; higher order terms are neglected. 
The resulting differential equation of motion is  
\begin{align}
\partial_t \vert \varphi_0(t) \rangle \approx \left( \partial_t \hat{\frak{h}}(t) \right) \vert \varphi_0(t) \rangle \mathrm{.}
\label{eqmo} 
\end{align}
Integration of Eq. (\ref{eqmo}) by the method of Magnus and Fer \cite{wilcox1967exponential} results to lowest order 
in 
\begin{align}
& \vert \varphi_0(t) \rangle \approx \exp \left( \hat{\frak{h}}(t) \right) \vert \varphi_0(t_0) \rangle \nonumber \\
& \hat{\frak{h}} = - \int \!\! d^3\mkern-2mu p \int_{t_0}^{t} \!\!\! dt^{\prime} \, \hat{c}_{\mathbf{p}}^{\dagger}(t^{\prime}) \int_{t_0}^{t^{\prime}} \!\!\! dt^{\prime \prime} \, 
\hat{c}_{\mathbf{p}}(t^{\prime \prime}) \mathrm{.}
\label{tldphi0} 
\end{align}
The next order term in the Magus Fer expansion,  
\begin{align}
\exp \left( \int_{t_0}^{t} \!\! dt^{\prime} \!\! \int_{t_0}^{t} \!\! dt^{\prime \prime} \left [ \partial_{t^{\prime}} \hat{\frak{h}}, \partial_{t^{\prime \prime}} \hat{\frak{h}} \right ] \right) 
\mathrm{,}
\label{magnusho}
\end{align}
is small and is neglected, see the discussion of the one- and two-photon operator terms in the next section. 

\subsection{Zero- one- and two-photon operator terms}

\noindent
The operator $\exp(\hat{\frak{h}}(t))$ is not unitary due to the coupling between ground state and continuum photon wavefunction. In order to isolate the unitary part of the wavefunction,  
$\hat{\frak{h}}(t)$ is split into an anti-Hermitian part plus small remainder. This is done here for zero-, one- and two-operator terms $\hat{\frak{h}} \approx \frak{h}^{(0)} + 
\hat{\frak{h}}^{(1)} + \hat{\frak{h}}^{(2)}$, respectively; higher order contributions and a phase term associated with the quantum statistics force \cite{even2023photon,xie2025maximal} are not considered here.

The expansion of $\hat{\frak{h}}$ is done, by first expanding $\hat{c}_{\mathbf{p}}$ in Eq. (\ref{dtldphip}) up to two-photon operator terms, $\hat{c}_{\mathbf{p}} \approx \hat{c}_{\mathbf{p}}^{(0)} 
+ \hat{c}_{\mathbf{p}}^{(1)} + \hat{c}_{\mathbf{p}}^{(2)}$ with
\begin{align}
& \hat{c}^{(0)}_{\mathbf{p}}(t) = \Omega(t) e^{i S(t)} \label{hatc} \\
& \hat{c}^{(1)}_{\mathbf{p}}(t) = \sum_{\kappa} \left( \hat{\Omega}_{\kappa}(t) - \Omega(t) \hat{\sigma}_{\kappa}(t) \right) e^{i S(t)} \nonumber \\
& \hat{c}^{(2)}_{\mathbf{p}}(t) = \sum_{\kappa, \kappa^{\prime}} \left( \frac{1}{2} \hat{\sigma}_{\kappa}(t) \hat{\sigma}_{\kappa^{\prime}}(t) - 
\hat{\Omega}_{\kappa}(t) \hat{\sigma}_{\kappa^{\prime}}(t) \right) e^{i S(t)} \mathrm{.}
\nonumber 
\end{align}

From this, $\hat{\frak{h}} = \hat{\frak{h}}^{(0)} + \hat{\frak{h}}^{(1)} + \hat{\frak{h}}^{(2)}$ in relation (\ref{tldphi0}) is determined up to second order with 
\begin{subequations}
\label{hath}
\begin{align}
& \hat{\frak{h}}^{(0)} = - \int \!\! d^3\mkern-2mu p \int_{t_0}^{t} \!\!\! dt^{\prime} \left( \hat{c}^{(0)*}_{\mathbf{p}} \hat{C}^{(0)}_{\mathbf{p}} \right)\!(t^{\prime}) 
\label{hath0} \\
& \hat{\frak{h}}^{(1)} = - \int \!\! d^3\mkern-2mu p \int_{t_0}^{t} \!\!\! dt^{\prime} \left( \hat{c}^{(0) *}_{\mathbf{p}} \hat{C}^{(1)}_{\mathbf{p}} + 
\hat{c}^{(1) \dagger}_{\mathbf{p}} \hat{C}^{(0)}_{\mathbf{p}} \right)\!(t^{\prime}) \label{hath1} \\
& \hat{\frak{h}}^{(2)} \!=\! - \! \int \!\! d^3\mkern-2mu p \!\! \int_{t_0}^{t} \!\!\! dt^{\prime} \! \left( \hat{c}^{(0) *}_{\mathbf{p}} \hat{C}^{(2)}_{\mathbf{p}} \!+\! 
\hat{c}^{(2) \dagger}_{\mathbf{p}} \hat{C}^{(0)}_{\mathbf{p}} \!+\! \hat{c}^{(1) \dagger}_{\mathbf{p}} \hat{C}^{(1)}_{\mathbf{p}} \right)\!(t^{\prime}) \mathrm{.}
\label{hath2}
\end{align}
\end{subequations}
Here, we have introduced $\hat{C}^{(j)}_{\mathbf{p}}(t) = \int_{-\infty}^{t} dt^{\prime} \hat{c}^{(j)}_{\mathbf{p}}(t^{\prime})$
($j=0,1,2$) for the inner time integral. Note that HHG terms should have the photon operator term outside of the inner time integral, 
i.e. they should all be represented by small letter terms $\hat{c}^{(1,2)}_{\mathbf{p}}$. This is also required to obtain unitary 
operators for HHG and QSHHG. To achieve this, we perform integration by parts of the terms in Eqs. (\ref{hath1}) and (\ref{hath2}) containing $\hat{C}^{(1,2)}_{\mathbf{p}}$. For example, $\int^{t} \! dt^{\prime} \hat{c}^{(0) *}_{\mathbf{p}} 
\hat{C}^{(1)}_{\mathbf{p}} = - \int^{t} \! dt^{\prime} \hat{c}^{(1)}_{\mathbf{p}} \hat{C}^{(0)*}_{\mathbf{p}} + R$, where $R$ is a 
small residual term. Finally the last term in Eq. (\ref{hath2}) has a one-photon operator in the ionization and emission part; as 
we are not interested in two color ionization, this term is neglected from hereon. 

Applying the above steps to Eq. ({\ref{hath}}) yields 
\begin{subequations}
\label{hathm}
\begin{align}
& \hat{\frak{h}}^{(0)} = - \int \!\! d^3\mkern-2mu p \int_{t_0}^{t} \!\!\! dt^{\prime} \left( \hat{c}^{(0)*}_{\mathbf{p}} \hat{C}^{(0)}_{\mathbf{p}} \right)\!(t^{\prime}) 
\label{hathm0} \\
& \hat{\frak{h}}^{(1)} \approx \int \!\! d^3\mkern-2mu p \int_{t_0}^{\infty} \!\!\! dt^{\prime} \left( \hat{c}^{(1)}_{\mathbf{p}} \hat{C}^{(0)*}_{\mathbf{p}} -
\hat{c}^{(1) \dagger}_{\mathbf{p}} \hat{C}^{(0)}_{\mathbf{p}} \right)\!(t^{\prime}) \label{hathm1} \\
& \hat{\frak{h}}^{(2)} \approx \int \!\! d^3\mkern-2mu p \int_{t_0}^{\infty} \!\!\! dt^{\prime} \left( \hat{c}^{(2)}_{\mathbf{p}} \hat{C}^{(0)*}_{\mathbf{p}} - 
\hat{c}^{(2) \dagger}_{\mathbf{p}} \hat{C}^{(0)}_{\mathbf{p}} \right)\!(t^{\prime}) \mathrm{,}
\label{hathm2}
\end{align}
\end{subequations}
where we also have let $t \rightarrow \infty$ in the HH and QSHH terms (\ref{hathm1}) and (\ref{hathm2}). 

Next, we evaluate the individual terms in Eq. (\ref{hathm}) by inserting Eq. (\ref{hatc}). For Eq. (\ref{hathm0}) one obtains 
\begin{align}
& \frak{h}^{(0)}(t) = - \int_{t_0}^{t} \!\!\! dt^{\prime} \gamma(t^{\prime}) = - \int_{t_0}^{t} \!\!\! dt^{\prime} \! \int \! d^3\mkern-2mu p \, \Gamma_{\mathbf{p}}(t^{\prime}) \nonumber \\
& \Gamma_{\mathbf{p}}(t) = \Omega^*(t)  e^{-i S(t) } \!\! \int_{t_0}^{t} \!\!\! dt^{\prime} \, \Omega(t^{\prime}) \, e^{i S(t^{\prime})} \mathrm{.}
\label{0op} 
\end{align}
Here, $\gamma(t)$ is a complex rate, and $\gamma + \gamma^*$ gives the optical field ionization rate from ground state to continuum. 

The one photon operator contribution (\ref{hathm1}) is worked out to be $\hat{\frak{h}}^{(1)} = \sum_{\kappa} \hat{\frak{h}}^{(1)}_{\kappa}$ with 
\begin{align}
\hat{\frak{h}}^{(1)}_{\kappa} = h_{\kappa} \hat{a}^{\dagger}_{\kappa} - h^{*}_{\kappa} \hat{a}_{\kappa} \mathrm{.}
\label{oneop1} 
\end{align}
The next order term in the Magnus Fer expansion (\ref{magnusho}) is a small phase term that does not contribute to HHG and can be neglected. Therefore, Eq. (\ref{tldphi0}) for HHG 
gives a coherent state with the HHG coefficient given by 
\begin{align}
h_{\kappa} & = - e^{-i \mathbf{k} \mathbf{x}_j} \!\! \int_{-\infty}^{\infty} \!\!\!\! dt^{\prime} e^{i \omega_k t^{\prime} } H_{k}(t^{\prime}) =  
\tilde{H}_{k} e^{-i \mathbf{k} \mathbf{x}_j} \label{hhg} \\
H_{k} & = \frac{\vert e \vert E_v}{\hbar} \Biggl\{ \bigl(\mathbf{e}_{\kappa} \mathbf{x}(t^{\prime}) \bigr)  
+ \int \!\! d^3\mkern-2mu p \, \overline{\sigma}_{\kappa}(t^{\prime}) \left( \Gamma_{\mathbf{p}}(t^{\prime}) + \mathrm{c.c} \right) \Biggr\} \mathrm{}
\nonumber 
\end{align}
with 
\begin{align}
& \mathbf{x}(t) = \int \!\! d^3\mkern-2mu p \, \mathbf{d}^*\!(\mathbf{p}_t) \, b_{\mathbf{p}}(t) + \mathrm{c.c.} 
\mathrm{,} \nonumber \\ 
& b_{\mathbf{p}}(t) = \int_{t_0}^{t} dt^{\prime} \Omega(t^{\prime}) \exp \left( i S(t^{\prime}, t) - \xi(t-t^{\prime}) \right) \mathrm{,}
\label{lewdip}
\end{align}
and $E_v = \sqrt{\hbar \omega_k / 2 \varepsilon_0 V}$ the vacuum electric field. The first term in Eq. (\ref{hhg}) agrees with previous work \cite{lewenstein1994theory}, where $\mathbf{x}(t)$ is the 
expectation value of the dipole moment driving HHG. Note that the subscript in $H_k$ is chosen $k$ to indicate that it does not depend on the HHG wavevector, but only on the frequency. It is also
assumed that harmonic and laser polarization vectors are parallel, so that the polarization index can be neglected as well. 

Usually dephasing is accounted for by a term $\xi(t-t^{\prime}) = (t-t^{\prime})/T_2$ with $T_2$ the dephasing time. Suppression of HHG grows with increasingly higher order electron returns. 
We use here a different filter which leaves the first return unfiltered and extinguishes all higher returns, 
\begin{align}
& \xi(t-t^{\prime}) = 0 \hspace{2cm} \, \mathrm{for} \,\, t-t^{\prime} \le \frac{T_0}{2} \nonumber \\ 
& \xi(t-t^{\prime}) = 10 (t-t^{\prime}) / T_0  \,\,\,\,\, \mathrm{for} \,\, \frac{T_0}{2} \le t-t^{\prime} \le T_0 \nonumber \\
& \xi(t-t^{\prime}) = \infty \hspace{1.9cm} \mathrm{for} \,\, t-t^{\prime} > T_0
\nonumber 
\end{align}
with optical period $T_0 = 2 \pi/ \omega_0$ and $\omega_0$ the laser frequency. The filter is confined to one optical cycle, as the associated convolution operation is numerically expensive. 

The QSHHG coefficients are determined from Eq. (\ref{hathm2}) as 
\begin{align}
\hat{\frak{h}}^{(2)} \!=\! \sum_{\kappa \kappa^{\prime}} f_{\kappa \kappa^{\prime}} \hat{a}^{\dagger}_{\kappa} \hat{a}^{\dagger}_{\kappa^{\prime}} 
\!-\! f^{*}_{\kappa \kappa^{\prime}} \hat{a}_{\kappa} \hat{a}_{\kappa^{\prime}} 
\!+\! g_{\kappa \kappa^{\prime}} \hat{a}^{\dagger}_{\kappa} \hat{a}_{\kappa^{\prime}} \!-\! g^{*}_{\kappa \kappa^{\prime}} \hat{a}_{\kappa} \hat{a}^{\dagger}_{\kappa^{\prime}} \mathrm{,}
\label{hatfrakh2_1}
\end{align}
where $f_{\kappa \kappa^{\prime}}$ is the coefficient for emission or absorption of two photons, 
\begin{align}
& f_{\kappa \kappa^{\prime}} = e^{-i (\mathbf{k} + \mathbf{k}^{\prime}) \mathbf{x}_j} \!\! \int_{-\infty}^{\infty} \!\!\!\! dt^{\prime} e^{i (\omega_k + \omega_{k^{\prime}}) t^{\prime}} \!\! 
\int \!\! d^3\mkern-2mu p \, \overline{\sigma}_{\kappa^{\prime}}(t^{\prime}) \times \nonumber \\
& \biggl\{ \frac{\vert e \vert E_{v}}{\hbar} (\mathbf{e}_{\kappa} \mathbf{x}_{\mathbf{p}}(t^{\prime})) 
+ \frac{1}{2} \overline{\sigma}_{\kappa}(t^{\prime}) \left[ \Gamma_{\mathbf{p}}(t^{\prime}) - \mathrm{c.c.} \right] \biggr\} \mathrm{.}
\label{f2ph} 
\end{align}
Emission of one and absorption of another photon is represented by 
\begin{align}
& g_{\kappa \kappa^{\prime}} = e^{-i (\mathbf{k} - \mathbf{k}^{\prime}) \mathbf{x}_j} \!\! \int_{-\infty}^{\infty} \!\!\!\! dt^{\prime} e^{i (\omega_k - \omega_{k^{\prime}}) t^{\prime}} \!\! 
\int \!\! d^3\mkern-2mu p \, \overline{\sigma}^*_{\kappa^{\prime}}(t^{\prime}) \times \nonumber \\
& \biggl\{ \frac{\vert e \vert E_{v}}{\hbar} (\mathbf{e}_{\kappa} \mathbf{x}_{\mathbf{p}}(t^{\prime})) 
+ \frac{1}{2} \overline{\sigma}_{\kappa}(t^{\prime}) \left[ \Gamma_{\mathbf{p}}(t^{\prime}) - \mathrm{c.c.} \right] \biggr\} 
\mathrm{,}
\label{g2ph} 
\end{align}
where 
\begin{align}
& \mathbf{x}_{\mathbf{p}}(t) = \mathbf{d}^*\!(\mathbf{p}_t) b_{\mathbf{p}}(t) - \mathrm{c.c.} 
\label{x-}
\end{align}
is the imaginary part of the ground state continuum transition dipole operator for a given $\mathbf{p}$. Here we have let $t \rightarrow \infty$ .

For the second order expansion term in Eq. (\ref{magnusho}) one has to keep finite $t$; it is worked out to be
\begin{align}
& \int_{t_0}^{t} \!\! dt^{\prime} \!\! \int_{t_0}^{t} \!\! dt^{\prime \prime} \left [ \partial_{t^{\prime}} \hat{\frak{h}}^{(2)}, \partial_{t^{\prime \prime}} \hat{\frak{h}}^{(2)} \right ] = 
\nonumber \\
& = \left( \partial_{t^{\prime}} f_{\kappa \kappa^{\prime}} \, \partial_{t^{\prime \prime}} f^{*}_{\kappa \kappa^{\prime}} - \mathrm{c.c.} \right)
\left(1 + \hat{n}_{\kappa} + \hat{n}_{\kappa^{\prime}} \right) \nonumber \\
& + \left( \partial_{t^{\prime}} g_{\kappa \kappa^{\prime}} \, \partial_{t^{\prime \prime}} g^{*}_{\kappa \kappa^{\prime}} - \mathrm{c.c.} \right) 
\left( \hat{n}_{\kappa^{\prime}} - \hat{n}_{\kappa} \right) \nonumber \\
& + \left( \partial_{t^{\prime}} g_{\kappa \kappa^{\prime}} \, \partial_{t^{\prime \prime}} f_{\kappa \kappa^{\prime}} 
- \partial_{t^{\prime}} f_{\kappa \kappa^{\prime}} \, \partial_{t^{\prime \prime}} g_{\kappa \kappa^{\prime}} \right) \hat{a}^{\dagger 2}_{\kappa} + \mathrm{h.c.} \nonumber \\
& + \left( \partial_{t^{\prime}} f_{\kappa \kappa^{\prime}} \, \partial_{t^{\prime \prime}} g^{*}_{\kappa \kappa^{\prime}} 
- \partial_{t^{\prime}} g^{*}_{\kappa \kappa^{\prime}} \, \partial_{t^{\prime \prime}} f_{\kappa \kappa^{\prime}} \right) \hat{a}^{\dagger 2}_{\kappa^{\prime}} + \mathrm{h.c.}
\label{magnusO2}
\end{align}
It only contains one-mode, two-photon terms and no mixed mode terms; all terms are second order in $f,g$; as $\vert f,g \vert \ll 1$, all terms in Eq. (\ref{magnusO2}) are neglected.

In the limit of a single perturbative mode, $\kappa^{\prime} = q$, the $\kappa^{\prime}$ sum can be dropped. Further, we write $f_{\kappa} = f_{\kappa q} + f_{q \kappa} \approx f_{\kappa q}$, 
and $g_{\kappa} = g_{\kappa q} + g_{q \kappa} \approx g_{\kappa q}$, as only the term $\kappa^{\prime} = q$ contributes significantly to QSHHG. This yields 
\begin{align}
f_{\kappa} & \!=\! e^{-i (\mathbf{k} + \mathbf{q}) \mathbf{x}_j} \!\!\! \int_{-\infty}^{\infty} \!\!\!\!\! dt^{\prime} e^{i (\omega_{k}+\omega_{q}) t^{\prime}} F_{k}(t^{\prime}) 
\! = \! \tilde{F}_{k} e^{-i (\mathbf{k} + \mathbf{q}) \mathbf{x}_j} \nonumber \\
F_{k} & \!=\! \left( \! \frac{\vert e \vert E_{v}}{\hbar} \! \right)^{\!2} \!\!\! \int \!\! d^3\mkern-2mu p \, \overline{\sigma}_{q}(t^{\prime}) 
\biggl\{ (\mathbf{e}_{\kappa} \mathbf{x}_{\mathbf{p}}(t^{\prime})) \!+\! i \overline{\sigma}_{\kappa}(t^{\prime}) \mathrm{Im}\left[ \Gamma_{\mathbf{p}}(t^{\prime}) \right] \biggr\} 
\label{f2ph} 
\end{align}
and  
\begin{align}
g_{\kappa} & \!=\! e^{-i (\mathbf{k} - \mathbf{q}) \mathbf{x}_j} \!\!\! \int_{-\infty}^{\infty} \!\!\!\!\! dt^{\prime} e^{i (\omega_{k}-\omega_{q}) t^{\prime}} G_{k}(t^{\prime}) 
\! = \! \tilde{G}_{k} e^{-i (\mathbf{k} - \mathbf{q}) \mathbf{x}_j} \nonumber \\
G_{k} & \!=\! \left( \! \frac{\vert e \vert E_{v}}{\hbar} \! \right)^{\!2} \!\!\! \int \!\! d^3\mkern-2mu p \, \overline{\sigma}^*_{q}(t^{\prime}) 
\biggl\{ (\mathbf{e}_{\kappa} \mathbf{x}_{\mathbf{p}}(t^{\prime})) \!+\! i \overline{\sigma}_{\kappa}(t^{\prime}) \mathrm{Im}\left[ \Gamma_{\mathbf{p}}(t^{\prime}) \right] \biggr\} \mathrm{,}
\label{g2ph} 
\end{align}
where $\omega_q$ and $\mathbf{q}$ are frequency and wavevector of the perturbative quantum light mode $q$. Frequencies $\omega_k$ and wavevector $\mathbf{k}$ continue to characterize the HHG and 
QSHHG modes. Again, like for HHG, the subscripts for $F_{k}$ and $G_{k}$ are chosen $k$ to indicate that they depend only on the QSHHG frequency or wavenumber. 

Combining all the results, we obtain the wavefuntion 
\begin{align}
& \left \vert \varphi_0 \right \rangle \! \approx \! \left \vert \varphi_{h} \right \rangle \left \vert \varphi_{m} \right \rangle \!=\! 
\hat{D}_h \hat{S}_m \left \vert \varphi_0(t_0) \right \rangle \mathrm{,}
\label{hhqc1}
\end{align}
where $t_0$ is the initial time, and the final time $t\rightarrow \infty$. The coherent state HHG operator is given by 
\begin{align}
\hat{D}_h = \exp \left( \sum_{\kappa} h_{\kappa} \hat{a}^{\dagger}_{\kappa} - h^*_{\kappa} \hat{a}_{\kappa} \right) \mathrm{,}
\label{hhq}
\end{align}
and the mixed mode, two-photon QSHHG operator is found to be 
\begin{align}
\hat{S}_{m}  = \exp \left( \sum_{\kappa} \left( f_{\kappa} \hat{a}^{\dagger}_{q} + g_{\kappa} \hat{a}_{q} \right) \hat{a}^{\dagger}_{\kappa} - \left( g^{*}_{\kappa} \hat{a}^{\dagger}_{q} 
+ f^{*}_{\kappa} \hat{a}_{q} \right) \hat{a}_{\kappa} \right) \mathrm{.}
\label{qshhq}
\end{align}
Here, we have assumed that HH and QSHH modes do not overlap. As a result, the wavefunction can be written as a product of the HH and QSHH time evolution operators. In case of mode overlap, 
the commutators between HH and QSHH operators need to be considered. 

In order to proceed, the above operators need to be applied to an initial state. Here, we use $\varphi_0(t_0) = \vert v \xi_q \rangle$ which consists of multi-mode vacuum states 
$\vert v \rangle = \prod_{\kappa} \vert v_{\kappa} \rangle$ for the harmonics, and a single BSV state for the perturbative mode, 
\begin{align}
\left \vert \xi_q \right \rangle = \hat{S}_q \left \vert v_q \right \rangle = e^{\frac{1}{2} (\xi \hat{a}_q^2
-\xi^* \hat{a}_q^{\dagger 2})} \left \vert v_q \right \rangle
\label{squvac}
\end{align}
with $\xi = r e^{i \theta}$. The normal ordered form of the squeezed state operator is given by \cite{fan2003operator}
\begin{align}
\hat{S}_q = \frac{1}{\sqrt{\cosh(r)}} e^{-\beta \hat{a}_q^{\dagger 2}} \, \colon \!\! e^{(\sech(r)-1) \hat{a}_q^{\dagger} \hat{a}_q} \colon e^{\beta^* \hat{a}_q^{2}}
\label{squn}
\end{align}
with $\beta = (1/2) \tanh(r) e^{-i \theta}$. Expressions written between colons are normal ordered.   

Finally, expectation values are dominantly determined by the ground state photon wavefunction; contributions of the continuum photon wavefunctions are negligible. As such, the expectation 
value of a general function of photon operators $\hat{O}(\ldots, \hat{a}_{\kappa}, \hat{a}^{\dagger}_{\kappa}, \ldots)$ is 
\begin{align}
& \langle \phi_0 \vert \hat{O} \vert \phi_0 \rangle = \langle \varphi_0 \vert \hat{D}_{\alpha}^{\dagger} \hat{U}_v^{\dagger} \hat{U}_i^{\dagger} \hat{O} \hat{U}_i \hat{U}_v \hat{D}_{\alpha} 
\vert \varphi_0 \rangle \nonumber \\
& = \langle \varphi_0 \vert \hat{O}(\ldots, \hat{a}_{\kappa} e^{-i \omega_k \tau}, \hat{a}^{\dagger}_{\kappa} e^{i \omega_k \tau}, \ldots) \vert \varphi_0 \rangle
\mathrm{.}
\label{expv}
\end{align}
The $U_v$ operator commutes with the other operators and cancels with its adjoint; the operator $D_{\alpha}$ adds the classical intense laser modes to the operators. As we are only interested in 
HH modes this is not explicitly written out. Finally, the operator $U_i$ makes the mode operators time dependent with $\tau = t - (\mathbf{n} \mathbf{x}_j) / c$.

\section{Phase matching}
\label{pm}

\subsection{High harmonic generation}
\label{pmhhg}

\noindent
So far we have developed the microscopic theory of HHG and QSHHG for a single atom. There is also a macroscopic aspect; when the radiation is summed over all atoms, propagation effects come into 
play. We first look at regular HHG and determine the macroscopic expectation value of the number operator $\hat{n} = \sum_{\kappa} \hat{n}_{\kappa}$,  
\begin{align}
& \langle \varphi_h \vert \hat{n} \vert \varphi_h \rangle = \sum_{\kappa} \! \left( \sum_j h^*_{\kappa} \right) \!\! 
\left( \sum_j h_{\kappa} \right) \nonumber \\
& = \frac{V N_0}{(2\pi)^3} \! \int \!\! d^3\mkern-1.5mu k \left( \int \!\! d^3\mkern-1.5mu x \, h^*_{\kappa} \! \right) \left( \int \!\! d^3\mkern-1.5mu x \, h_{\kappa} \! \right) 
\mathrm{,}
\label{expecthh}
\end{align}

While the integrals in Eq. (\ref{expecthh}) can be evaluated numerically, we give here an order of magnitude estimate. We assume that all dominant laser and harmonic modes propagate predominantly 
along $z$ and are polarized along $x$. 
Also, it is assumed that the transverse profile of the pump laser remains approximately unchanged, i.e. independent of $z$. As a result, $\tilde{H}_{k}$ can be factorized in transverse ($x, y$) and longitudinal ($z$) parts. 

First, the integral of each $h_{\kappa}$ over the transverse spatial coordinates is performed; in the spirit of saddle point integration, the exponent of $h_{\kappa}$ is expanded up to second order 
in $x,y$, and $x,y$ are set to zero in the pre-exponential terms. Then, 
\begin{align}
& \int_{-\infty}^{\infty} \!\!\!\!\! d^3\mathbf{x} \, h_{\kappa}(\omega,\mathbf{x}) \! \approx \!\!\! \int_{0}^{l_i} \!\!\!\! dz h_{\kappa}(\omega,z) \!\! \int_{-\infty}^{\infty} \!\!\!\!\! dx dy \, 
e^{-\frac{x^2 \!+\! y^2}{2\mathrm{w}_{k}^2}} \! e^{-i (k_x x \!+\! k_y y)}
\nonumber \\
& = 2 \pi \mathrm{w}_{k}^2 e^{-\frac{\mathrm{w}_{k}^2}{2}(k_x^2 + k_y^2)} \int_{0}^{l_i} \!\!\!\! dz h_{\kappa}(\omega,z) \mathrm{,}
\label{htransx}
\end{align}
where $l_i$ is the interaction length and $\mathrm{w}_{k} = \mathrm{w}(\omega_{k})$ is the width of the transverse beam profile of harmonic mode $\kappa$. Note that due to the nonlinearity of the
ionization process $\mathrm{w}_{k} / \mathrm{w}_0 < 1$ with $\mathrm{w}_0 = \mathrm{w}(\omega_0)$. 

Second, the integral over $dk_x dk_y$ is performed. As harmonic emission is mainly directed along the laser propagation direction, one can approximate $k_z$ in the plane wave of Eq. (\ref{hhg})
by $k_z^2 = k^2 - (k_x^2 + k_y^2) \approx k^2$ with $k = \omega_k/c$; consequently also $d k_z \approx dk$ in Eq. (\ref{expecthh}). This corresponds to neglecting the z-dependent evolution of the
transverse Gaussian profile of the harmonic pulse. Thus, the $dk_x dk_y$ integral in Eq. (\ref{expecthh}) can be performed separately as well, 
\begin{align}
\int_{-\infty}^{\infty} \!\!\!\! dk_x dk_y e^{-\mathrm{w}_{k}^2(k_x^2 + k_y^2)} = \frac{\pi}{\mathrm{w}_{k}^2}
\label{htransk}
\end{align}

Third, the $z$-integral in Eq. (\ref{htransx}) is $\int_0^{l_i} dz \exp(-i(k - N k_0)z) = \int_0^{l_i} dz \exp(-i \Delta k z) $,
where $\Delta k$ is the difference between the harmonic wavevector and $N$ times the fundamental wavevector. Here, we have taken 
$\kappa$ as the mode with frequency $\omega_k = N \omega_0$. There are several sources that change $k_0$ and result in dephasing; 
beam geometry, gas refractive index, and the refractive index of the free electrons. The absolute square of the integral has a maximum at $\Delta k l_i = \pi$ \cite{boyd2020nonlinear}. We assume that $l_i < \pi / \Delta k$, so that $\int_0^{l_i} dz 
\exp(-i \Delta k z) \approx l_i$.

Finally, the $d\omega_k$ integral drops out, when looking at the differential expectation value of the number operator. Inserting the above results in Eq. (\ref{expecthh}) yields 
\begin{align}
\frac{d \langle \hat{n} \rangle}{d \omega_k} & = \frac{\langle \varphi_h \vert \hat{n} \vert \varphi_h \rangle}{d \omega_k} = c_{k}^2 \, \vert \tilde{H}_{k}(\omega_k) \vert^2 
\nonumber \\
c_{k}^2 & = \frac{ \left( N_0 \mathrm{w}_{k} l_i \right)^2}{2c} V \mathrm{,} 
\label{expecthh1}
\end{align}
where the atomic position coordinate is now set to $\mathbf{x} = 0$ in $\tilde{H}_{k}$. The quantization volume cancels out, as the single atom response $\tilde{H}_{k} \propto 1/\sqrt{V}$. 
As the transverse wavevector modes have been integrated over, the modes are counted only by frequency $\omega_k$ or wavenumber $k$. We also define the number of photons emitted into a single mode 
with wavenumber $k$, 
\begin{align}
\langle \hat{n} \rangle_{k} = \frac{d \langle \hat{n} \rangle}{d \omega_k} \Delta \omega 
\label{expecthh1mode}
\end{align}
with $\Delta \omega$ the harmonic mode width. The number of photons emitted into one harmonic interval $\omega_0$ about harmonic order $N$ is given by 
\begin{align}
\langle \hat{n} \rangle_{\!N} = c_{k}^2 \int_{(N-\frac{1}{2})\omega_0}^{(N+\frac{1}{2})\omega_0} \!\!\!\! d \omega_k 
\vert \tilde{H}_{k}(\omega_k) \vert^2 = \vert h_{N} \vert^2
\mathrm{,}
\label{expecthh2}
\end{align}
In summary, $\vert h_{N} \vert^2$ determines the macroscopic emission of HH photons in all transverse spatial modes in the frequency interval $(N-\frac{1}{2}) \omega_0 \le \omega_k \le 
(N+\frac{1}{2}) \omega_0$.

\subsection{Quantum sideband high harmonic generation}
\label{pmqshhg}

\noindent
Next we look at phase matching of QSHHG. First the expecation value of the number operator needs to be evaluated 
by using the Bogoliubov transformation \cite{gerry2023introductory}, 
\begin{align}
\left \langle \varphi_m \right \vert \hat{n}_{\kappa} \left \vert \varphi_m \right \rangle = 
\left \langle \xi_q  v_{\kappa} \right \vert \hat{S}_m^{\dagger} \hat{a}^{\dagger}_{\kappa} \hat{S}_m 
\hat{S}_m^{\dagger} \hat{a}_{\kappa} \hat{S}_m \left \vert v_{\kappa} \xi_{q} \right \rangle \mathrm{.}
\label{bogi}
\end{align}
The transformation of the annihilation operator is found to be 
\begin{align}
\hat{S}_m^{\dagger} \hat{a}_{\kappa} \hat{S}_m = \cosh(M_{\kappa}) \hat{a}_{\kappa} + \frac{\sinh(M_{\kappa})}{M_{\kappa}} \left( f_{\kappa} a_{q}^{\dagger} + g_{\kappa} \hat{a}_q \right)
\label{bogi1}
\end{align}
with $M_{\kappa}^2 = \vert f_{\kappa} \vert^2 - \vert g_{\kappa} \vert^2$. Using Eqs. (\ref{bogi1}) in (\ref{bogi}), followed by another Bogoliubov transformation in $\hat{S}_q$, yields 
\begin{align}
& \langle \varphi_m \vert \hat{n}_{\kappa} \vert \varphi_m \rangle \approx \cosh^2(r) \left \vert Z_{\kappa} \right \vert^2 \nonumber \\ 
& Z_{\kappa} = f_{\kappa} - g_{\kappa} \tanh(r) e^{i \theta} \mathrm{.}
\label{ninf}
\end{align}
Here, $M_{\kappa} \ll 1$ and $r \gg 1$ has been used. 

To obtain the macroscopic expectation value, the same procedure is followed as for HHG, 
\begin{align}
& \left \langle \varphi_m \left \vert \hat{n} \right \vert \varphi_m \right \rangle 
= N_0^2 \cosh^2(r) \sum_{\kappa q} \left \vert \int_{V} \!\! d^3\mkern-2.5mu x \, Z_{\kappa} \right \vert^2 
\nonumber \\ 
&  \approx \frac{V^2 N_0^2 (\Delta q)^3 \cosh^2(r)}{(2\pi)^6}  \! \int \!\! d^3\mkern-1.5mu k \left \vert \int_{V} d^3\mkern-2.5mu x \, Z_{\kappa} \right \vert^2 \mathrm{,}
\label{nf}
\end{align}
where we have converted the sum over $\kappa$ into an integral as before. Further, the mode volume of the perturbative mode $q$ is introduced as $(\Delta q)^3 = \Delta q_x \Delta q_y \Delta q_z 
= (2 \pi)^3/V$. Like before, the exponent is Taylor expanded with regard to the transverse space variables, and $x, y$ is set to zero in the preexponent; the factured out $dx dy$ integrals 
for $f_{\kappa}$ and $g_{\kappa}$ are 
\begin{align}
& \int_{-\infty}^{\infty} \!\!\! dx dy \, e^{-\frac{x^2+y^2}{2\mathrm{w}_{k}^2}} e^{-i ((k_x \pm q_x) x + (k_y \pm q_y) y)} = 
\nonumber \\
& 2 \pi \mathrm{w}_{k}^{2} e^{-\frac{\mathrm{w}_{k}^2}{2}((k_x \pm q_x)^2 + (k_y \pm q_y)^2)}
\mathrm{.}
\label{nftrv2}
\end{align}
As harmonic emission is mainly directed along the laser propagation direction, again $k_z \approx k = \omega_k/c$ and $dk_z \approx dk = d\omega_k/c$. As before, the $dk_x dk_y$ integral over 
(\ref{nftrv2}) squared can be separated out and performed as well to give 
\begin{align}
\int_{-\infty}^{\infty} dk_x dk_y e^{-\mathrm{w}_{k}^2((k_x \pm q_x)^2 + (k_y \pm q_y)^2)} = \frac{\pi}{\mathrm{w}_{k}^2} \mathrm{;}
\label{transi}
\end{align}
the $q_x, q_y$ dependence vanishes. Assuming that the interaction length is shorter than the dephasing length, the integral over $dz$ gives a factor $l_i^2$, as before. 

After the transverse wavevectors have been integrated out, the modes are identified by the QSHHG frequency (wavenumber) alone. Inserting the above results in Eq. (\ref{nf}) determines the number 
of photons emitted in all modes. The number of photons emitted into a single mode $k$ is 
\begin{align}
& \langle \hat{n} \rangle_{k} \!=\! c_q^2 \cosh^2(r) \left \vert \zeta_{k} \right \vert^2 \Delta \omega \nonumber \\
& \vert \zeta_{k} \vert^2 = \left \vert \tilde{F}_{k} - \tilde{G}_{k} \tanh(r) e^{i \theta} \right \vert^2 \mathrm{}
\label{nexpvf}
\end{align}
with 
\begin{align}
c_q^2 = \frac{ \left( N_0 \mathrm{w}_{k} l_i \right)^2}{2c} V^2 \frac{(\Delta q)^3}{(2 \pi)^3} \mathrm{.}
\label{cqsq}
\end{align}
The steps leading from Eq. (\ref{ninf}) to (\ref{nexpvf}) are summarized in the following. In the transition from $Z_{\kappa}$ to $\zeta_{k}$ the identity $V^2 / (2\pi)^6 \Delta^3 k \Delta^3 q = 1$
is used and the limit from discrete to continuous modes is performed. Whereas, $Z_{\kappa}$ contains the sum over all atoms and represents a single mode $\kappa = (\mathbf{k},s)$, $\zeta_{k}$ stands 
for a single mode $k$, as the transverse QSHHG wavevectors have been integrated out. Further, the sum over all atoms has been evaluated so that $\zeta_{\kappa}$ does no longer depend on 
$\mathbf{x}_j$. Finally, the factor $V^2 \left \vert \zeta_{\kappa} \right \vert^2$ is independent of the quantization volume. 

Comparison of Eq. (\ref{nexpvf}) with (\ref{expecthh1mode}) shows that the ratio of propagation related pre-factors of QSHHG to HHG is $1/(2\pi)^3 (\Delta q)^3$. The mode volume is determined by the 
beam parameters of the quantum field spectral width $\Delta \lambda_q$ and transverse beam width $\mathrm{w}_q$, from which one obtains $\Delta q_x = \Delta q_y \approx (2 \pi) / \mathrm{w}_q$ and 
$\Delta q_z = 2 \pi \Delta \lambda_q / \lambda_q^2$. As a result, $(\Delta q)^3 / (2\pi)^3 = \Delta \lambda_q / (\mathrm{w}_q^2 \lambda_q^2)$. This is an approximation. In reality the
single BSV mode is composed of a number of plane wave modes. This fact has been accounted for by replacing the mode volume of a single plane wave mode with the mode volume of the BSV mode. The quality of this assumption is corroborated by the good agreement between experiment and theory with regard to the number of emitted sideband photons, see the 
manuscript. 

Finally, like before we define QSHHG in the frequency interval $(N-\frac{1}{2}) \omega_0 \le \omega_k \le (N+\frac{1}{2}) \omega_0$ and in all spatial modes as 
\begin{align}
\langle  \hat{n} \rangle_{\!N} & \!=\! c_q^2 \cosh^2(r) \int_{(N-\frac{1}{2})\omega_0}^{(N+\frac{1}{2})\omega_0} \!\!\!\! d \omega_k \left \vert \zeta_{k} \right \vert^2 
= \cosh^2(r) \left \vert \zeta_{N} \right \vert^2 \mathrm{.}
\label{nqsom0}
\end{align}

\section{Properties of QSHHG}

\subsection{Normal ordered wavefunction}

\noindent
To characterize the quantum properties of QSHHG we need a simpler expression for the wavefunction. This can be achieved in the limit of intense quantum fields perturbing HHG. We proceed by swapping the order of operators in Eq. (\ref{hhqc1}), 
\begin{align}
& \left \vert \varphi_m(t) \right \rangle \!=\! \hat{S}_m \left \vert v_{\kappa} \xi_q \right \rangle = \hat{S}_q \hat{S}_q^{\dagger} \hat{S}_m \hat{S}_q 
\left \vert v_{\kappa} v_q \right \rangle = \hat{S}_q \hat{S}^{\prime}_m \left \vert v v_q \right \rangle \mathrm{.}
\label{swap}
\end{align}
The transformed operator $\hat{S}^{\prime}_m$ is evaluated by Taylor expansion of $\hat{S}_m$ and by calculating the Bogoliubov transformation with regard to $\hat{S}_q$; this amounts to 
\begin{align}
\hat{S}^{\prime}_{m}  = \exp \left( \sum_{\kappa} \left( f^{\prime}_{\kappa} \hat{a}^{\dagger}_{q} + g^{\prime}_{\kappa} \hat{a}_{q} \right) \hat{a}^{\dagger}_{\kappa} - 
\left( g^{\prime *}_{\kappa} \hat{a}^{\dagger}_{q} + f^{\prime *}_{\kappa} \hat{a}_{q} \right) \hat{a}_{\kappa} \right) \mathrm{} 
\label{Sm}
\end{align}
with 
\begin{align}
& f^{\prime}_{\kappa} = f_{\kappa} \cosh(r) - g_{\kappa} \sinh(r) e^{i \theta} = \cosh(r) Z_{\kappa} \nonumber \\
& g^{\prime}_{\kappa} = - f_{\kappa} \sinh(r) e^{-i \theta} + g_{\kappa} \cosh(r) \mathrm{.}
\label{tildfg}
\end{align}
The swap in Eq. (\ref{swap}) is important, as $f_{\kappa}, g_{\kappa} \ll 1$ are small and $r$ is big. As such, an exponential operator of the order of unity acts on a very large exponential 
operator in the original expression. This would make it difficult to isolate the leading order terms in $\hat{S}_m$, and important terms can be easily lost. 

Next, the exponent of the sum over $\kappa$ modes in Eq. (\ref{Sm}) is split into a product of single mode exponents by using the Baker-Campbell-Haussdorff (BCH) formula, $e^{\hat{A}+\hat{B}} = 
e^{\hat{A}} e^{\hat{B}} e^{-(1/2)[\hat{A},\hat{B}]} \cdots$; we only evaluate the lowest order commutator. Defining $\hat{J}^{\dagger}_{\kappa q} = f^{\prime}_{\kappa} \hat{a}^{\dagger}_{q} + 
g^{\prime}_{\kappa} \hat{a}_{q}$ and assuming two modes $j = \kappa, \kappa^{\prime}$ yields 
\begin{align}
e^{ \sum_{j = \kappa, \kappa^{\prime}} \! \hat{J}^{\dagger}_{jq} \hat{a}_j^{\dagger} - \hat{J}_{jq} \hat{a}_j } = 
\hat{B} \prod_{j = \kappa, \kappa^{\prime}} \exp \left( \hat{J}^{\dagger}_{jq} \hat{a}_j^{\dagger} - \hat{J}_{jq} \hat{a}_j \right) \mathrm{,}
\label{Smprod}
\end{align}
where 
\begin{align}
& \hat{B} = \exp \left( -(1/2) \left[ \hat{J}^{\dagger}_{\kappa q} \hat{a}_{\kappa}^{\dagger} - \hat{J}_{\kappa q} \hat{a}_{\kappa},  
\hat{J}^{\dagger}_{\kappa^{\prime} q} \hat{a}_{\kappa^{\prime}}^{\dagger} - \hat{J}_{\kappa^{\prime} q} \hat{a}_{\kappa^{\prime}} \right] \right) \nonumber \\
& = \exp \Bigl( -(1/2) \Bigl\{ \hat{a}^{\dagger}_{\kappa} \hat{a}^{\dagger}_{\kappa^{\prime}} \left[ \hat{J}^{\dagger}_{\kappa q}, \hat{J}^{\dagger}_{\kappa^{\prime} q} \right] 
+ \hat{a}_{\kappa} \hat{a}_{\kappa^{\prime}} \left[ \hat{J}_{\kappa q}, \hat{J}_{\kappa^{\prime} q} \right] \nonumber \\
& + \left(\hat{a}^{\dagger}_{\kappa} \hat{a}_{\kappa^{\prime}} - \hat{a}^{\dagger}_{\kappa^{\prime}} \hat{a}_{\kappa} \right) \left[ \hat{J}^{\dagger}_{\kappa q}, \hat{J}_{\kappa^{\prime} q} \right]
\Bigr\} \Bigr)
\label{Smprod1}
\end{align}
As 
\begin{align}
& \left[ \hat{J}^{\dagger}_{\kappa q}, \hat{J}_{\kappa^{\prime} q} \right] = - \left(f^{\prime}_{\kappa} f^{\prime*}_{\kappa^{\prime}} - g^{\prime}_{\kappa} g^{\prime*}_{\kappa^{\prime}} \right)
= - \left(f_{\kappa} f^*_{\kappa^{\prime}} - g_{\kappa} g^*_{\kappa^{\prime}} \right) \ll 1 \nonumber \\
& \left[ \hat{J}^{\dagger}_{\kappa q}, \hat{J}^{\dagger}_{\kappa^{\prime} q} \right] = - \left( f^{\prime}_{\kappa} g^{\prime}_{\kappa^{\prime}} - f^{\prime}_{\kappa^{\prime}} 
g^{\prime}_{\kappa} \right) = - \left( f_{\kappa} g_{\kappa^{\prime}} - f_{\kappa^{\prime}} g_{\kappa} \right) \ll 1 \nonumber \\
& \left[ \hat{J}_{\kappa q}, \hat{J}_{\kappa^{\prime} q} \right] = \left( f^{\prime*}_{\kappa} g^{\prime*}_{\kappa^{\prime}} - f^{\prime*}_{\kappa^{\prime}} 
g^{\prime*}_{\kappa}\right) = \left( f^*_{\kappa} g^*_{\kappa^{\prime}} - f^*_{\kappa^{\prime}} g^*_{\kappa} \right) \ll 1
\mathrm{,}
\label{Smprod2}
\end{align}
the commutator $ \hat{B} \approx 1$ is negligible; and so are all higher order commutator terms.  

In order to simplify $\hat{S}^{\prime}_{m}$, it is normal ordered, i.e. all destruction operators are moved to the right of the creation operators. As a result, the destruction operators act 
directly on the vacuum states and drop out. Normal ordering results in \cite{agrawal1977ordering}
\begin{align}
& \hat{S}^{\prime}_m = \prod_{\kappa} \mathcal{N}_{\kappa} 
\exp \left( \frac{1}{2} q_{\kappa} f^{\prime}_{\kappa} \! \left( g^{\prime}_{\kappa} \hat{a}_{\kappa}^{\dagger 2} - g^{\prime*}_{\kappa} \hat{a}_{q}^{\dagger 2} \right) 
\!+\! q^{\prime \prime}_{\kappa} f^{\prime}_{\kappa} \hat{a}_{\kappa}^{\dagger} \hat{a}_{q}^{\dagger} \right) 
\nonumber \\
& \colon \! \exp \left(-2 q^{\prime}_{\kappa} \left(\hat{a}_{\kappa}^{\dagger} \hat{a}_{\kappa} + \hat{a}_{q}^{\dagger} \hat{a}_{q} \right)
\!+\! q^{\prime \prime \prime}_{\kappa} \left( g^{\prime}_{\kappa} \hat{a}_{\kappa}^{\dagger} \hat{a}_{q} - g^{\prime *}_{\kappa} \hat{a}_{\kappa} \hat{a}_{q}^{\dagger} \right) 
\right) \! \colon
\nonumber \\
& \exp \left( \frac{1}{2} q_{\kappa} \! \left( f^{\prime *}_{\kappa} g^{\prime *}_{\kappa}  \hat{a}_{\kappa}^{ 2} - f^{\prime *}_{\kappa} g^{\prime}_{\kappa} \hat{a}_{q}^{2} 
\right) - q^{\prime \prime}_{\kappa} f^{\prime *}_{\kappa} \hat{a}_{\kappa} \hat{a}_q \right) \mathrm{.}
\label{nordS1}
\end{align}
The normalization factor is given by 
\begin{align}
 \frak{N}_{\kappa} = & \frac{1/\cosh^2(m_{\kappa})}{\sqrt{1+2 \left( \vert f^{\prime}_{\kappa} \vert^2 + \vert g^{\prime}_{\kappa} \vert^2 \right)
\left( \displaystyle {\frac{\tanh(m_{\kappa})}{m_{\kappa}} } \right)^2  + \tanh^4(m_{\kappa}) }} \nonumber \\
& \approx \frac{1}{\sqrt{1 + \vert f^{\prime}_{\kappa} \vert^2}} \mathrm{,}
\label{norm}
\end{align}
where $4 m^{2}_{\kappa} = \vert f^{\prime}_{\kappa} \vert^2 - \vert g^{\prime}_{\kappa} \vert^2 = \vert f_{\kappa} \vert^2 - \vert g_{\kappa} \vert^2 \ll 1$. The other parameters are 
\begin{align}
& q_{\kappa} = \frac{\tanh^2(m_{\kappa})}{m^{2}_{\kappa} \left( 1+ \tanh^2(m_{\kappa}) \right)^2 + \vert g^{\prime}_{\kappa} \vert^2 
\tanh^2(m_{\kappa}) } \approx \frac{1}{1 + \vert f^{\prime}_{\kappa} \vert^2} \nonumber \\
& q^{\prime}_{\kappa} = q_{\kappa} \left( \frac{1}{4} \left( \vert f^{\prime}_{\kappa} \vert^2 + \vert g^{\prime}_{\kappa} \vert^2 \right) + 
m^{2}_{\kappa} \tanh^2(m^{2}_{\kappa}) 
\right) \nonumber \\
& q^{\prime \prime}_{\kappa} = \frac{m_{\kappa} \tanh(m_{\kappa}) \left(1 + \tanh^2(m_{\kappa}) \right) }
{m^{2}_{\kappa} \left( 1+ \tanh^2(m_{\kappa}) \right)^2 + \vert g^{\prime}_{\kappa} \vert^2 \tanh^2(m_{\kappa}) } 
\approx \frac{1}{1 + \vert f^{\prime}_{\kappa} \vert^2} \nonumber \\
& q^{\prime \prime \prime}_{\kappa} = \frac{m_{\kappa} \tanh(m_{\kappa}) \left(1 - \tanh^2(m_{\kappa}) \right) }
{m^{2}_{\kappa} \left( 1+ \tanh^2(m_{\kappa}) \right)^2 + \vert g^{\prime}_{\kappa} \vert^2 \tanh^2(m_{\kappa}) } \nonumber \\
\label{q}
\end{align}
The second line of Eq. (\ref{norm}) and the last, approximate expressions in the first and third line of Eq. (\ref{q}) are valid for $r \gg 1$ and $f_{\kappa},g_{\kappa} \ll 1$. 

We first proceed with a single harmonic mode and then generalize the result to multi harmonic modes. Inserting Eq. (\ref{nordS1}) in (\ref{swap}), keeping only a single harmonic mode, and using the
normal ordered form (\ref{squn}) of $\hat{S}_q$ yields the simpler expression 
\begin{align}
& \left \vert \varphi_m \right \rangle_{\kappa} \!=\! \frac{\frak{N}_{\kappa}}{\sqrt{\cosh(r)}} e^{\frac{1}{2} q_{\kappa} f^{\prime}_{\kappa} g^{\prime}_{\kappa} 
\hat{a}_{\kappa}^{\dagger 2} } e^{-\beta \hat{a}_q^{\dagger 2}} \colon \! e^{(\sech(r)-1) \hat{a}_q^{\dagger} \hat{a}_q} \! \colon \nonumber \\
& e^{\beta^* \hat{a}_q^{2}} e^{- \frac{1}{2} q_{\kappa} f^{\prime}_{\kappa} g^{\prime *}_{\kappa} \hat{a}_{q}^{\dagger 2}}
e^{q^{\prime \prime}_{\kappa} f^{\prime}_{\kappa} \hat{a}_{\kappa}^{\dagger} \hat{a}_{q}^{\dagger}} \left \vert v_{\kappa} v_q \right \rangle \mathrm{,}
\label{wfno}
\end{align}
where all exponents containing annihilation operators act on the vacuum and have become unity. The operators from $\hat{S}_q$ still need to be normal ordered. This is done using the IWOP 
method \cite{qiu2022coherent,fan2003operator} and yields 
\begin{align}
\left \vert \varphi_m \right \rangle_{\kappa} \!=\! 
\frac{\frak{N}_{\kappa} e^{Q_1 \hat{a}_{\kappa}^{\dagger 2} } e^{-Q_2 \hat{a}_{q}^{\dagger 2} } e^{Q_m \hat{a}_{\kappa}^{\dagger} \hat{a}_{q}^{\dagger}}}
{\sqrt{\cosh(r)} \sqrt{1+2 \beta^* q_{\kappa} f^{\prime}_{\kappa} g^{\prime *}_{\kappa}} } 
\left \vert v_{\kappa} v_q \right \rangle \mathrm{,}
\label{wfno1}
\end{align}
where 
\begin{align}
& Q_1 = \frac{1}{2} q_{\kappa} f^{\prime}_{\kappa} g^{\prime}_{\kappa} \!+\! 
\frac{\beta^* \left( q^{\prime \prime}_{\kappa} f^{\prime}_{\kappa} \right)^2}{1+2 \beta^* q_{\kappa} f^{\prime}_{\kappa} g^{\prime *}_{\kappa}} \approx 0 \nonumber \\
& Q_2 = \beta \!+\! \frac{\sech^2(r) q_{\kappa} f^{\prime}_{\kappa} g^{\prime *}_{\kappa}}{2 \left( 1+2 \beta^* q_{\kappa} f^{\prime}_{\kappa} g^{\prime *}_{\kappa} \right)} 
\approx \frac{\beta}{1 + \vert Z_{\kappa} \vert^2} \nonumber \\
& Q_m = \frac{\sech(r) q^{\prime \prime}_{\kappa} f^{\prime}_{\kappa} }{1+2 \beta^* q_{\kappa} f^{\prime}_{\kappa} g^{\prime *}_{\kappa} } \approx Z_{\kappa} \mathrm{.}
\label{Qx}
\end{align}
The approximate expressions were obtained by using $r \gg 1$ and $f_{\kappa},g_{\kappa} \ll 1$, $\tanh(r) \approx 1$, Eq. (\ref{q}), and  
\begin{align}
1 + 2 \beta^* q_{\kappa} f^{\prime}_{\kappa} g^{\prime *}_{\kappa} \approx \frac{1 + \vert Z_{\kappa} \vert^2}{1 + \vert f^{\prime}_{\kappa} \vert^2} \mathrm{.}
\label{nopref}
\end{align}
The last relation is also used in the evaluation of the pre-exponential factor in Eq. (\ref{wfno1}). Inserting the above approximations in Eq. (\ref{wfno1}) yields the final expression for the
QSHHG wavefunction in the limit of a bright quantum perturbation, 
\begin{align}
& \left \vert \varphi_m \right \rangle_{\kappa} \!=\! \frac{\frak{N}_{\kappa}}{\sqrt{\cosh(r)}} \exp\left(Z_{\kappa} \hat{a}_{\kappa}^{\dagger} 
\hat{a}_{q}^{\dagger} \right) \exp \left(- \beta_{\kappa} \hat{a}_q^{\dagger 2} \right) \left \vert v_{\kappa} v_2 \right \rangle
\nonumber \\
& \beta_{\kappa} = \beta / (1 + \vert Z_{\kappa} \vert^2) \,\,\,\,\,\,\, \frak{N}_{\kappa} \approx \frac{1}{\sqrt{1 + \vert Z_{\kappa} \vert^2} } \mathrm{.}
\label{wfno2}
\end{align}
The approximate wavefunction (\ref{wfno2}) can be shown to be normalized, accurate to first order in $\vert Z_{\kappa} \vert^2 \ll 1$. 

In the multi harmonic mode case we start again from Eq. (\ref{nordS1}); each $\kappa, q$ mixed exponent in $\hat{S}_m$ contains $\hat{a}_q^{\dagger}$, $\hat{a}_q$ terms that need to be normal 
ordered. The $\hat{a}_{\kappa}^{\dagger}$ and $\hat{a}_{\kappa}$ operators of different harmonic modes commute; therefore all terms containing only harmonic operators with $\hat{a}_{\kappa}$ 
in it commute to the vauum state and drop out. Inserting Eq. (\ref{nordS1}) in 
(\ref{swap}) yields 
\begin{align}
& \left \vert \varphi_m \right \rangle \!=\! \prod_{\kappa} \frac{\frak{N}_{\kappa}}{\sqrt{\cosh(r)}} e^{\frac{1}{2} q_{\kappa} f^{\prime}_{\kappa} g^{\prime}_{\kappa} 
\hat{a}_{\kappa}^{\dagger 2} } e^{-\beta \hat{a}_q^{\dagger 2}} \colon \! e^{(\sech(r)-1) \hat{a}_q^{\dagger} \hat{a}_q} \! \colon \nonumber \\
& \times e^{\beta^* \hat{a}_q^{2}} \, e^{- \frac{1}{2} q_{\kappa} f^{\prime}_{\kappa} g^{\prime *}_{\kappa} \hat{a}_{q}^{\dagger 2}}
e^{q^{\prime \prime}_{\kappa} f^{\prime}_{\kappa} \hat{a}_{\kappa}^{\dagger} \hat{a}_{q}^{\dagger}} \, \colon \!e^{-2 q^{\prime}_{\kappa} \hat{a}_{q}^{\dagger} \hat{a}_q} \! \colon 
e^{q^{\prime \prime \prime}_{\kappa} g^{\prime}_{\kappa} \hat{a}^{\dagger}_{\kappa} \hat{a}_q} 
\nonumber \\
& \times e^{-\frac{1}{2} q_{\kappa} f^{\prime *}_{\kappa} g^{\prime}_{\kappa}\hat{a}_q^2} \, \left \vert v_{\kappa} v_q \right \rangle \mathrm{,}
\label{wfnomm}
\end{align}
where in addition to the single mode case above, the last three terms containing $\hat{a}_q$ in Eq. (\ref{wfnomm}) appear in each harmonic mode and need to be normal ordered. Luckily, normal 
ordering of these terms gives contributions of higher order which can be neglected. Therefore, the multi-mode expression looks similar to the single mode one, 
\begin{align}
& \left \vert \varphi_m \right \rangle \!=\! \prod_{\kappa} \frac{\overline{\frak{N}}}{\sqrt{\cosh (r)}} \exp \left( Z_{\kappa} \hat{a}_{\kappa}^{\dagger} 
\hat{a}_{q}^{\dagger} \right) \exp \left(- \overline{\beta} \hat{a}_q^{\dagger 2} \right) \left \vert v_{\kappa} v_q \right \rangle \nonumber \\
& \overline{\beta} \approx \frac{\beta}{1 + \sum_{\kappa} \vert Z_{\kappa} \vert^2} \mathrm{,} \,\,\,\,\,\,\, \overline{\frak{N}} \approx \frac{1}{\sqrt{1 + \sum_{\kappa} \vert Z_{\kappa} \vert^2}} 
\mathrm{.}
\label{wfno2mm}
\end{align}
In the last equation, $\prod_{\kappa} 1/(1 + \vert Z_{\kappa} \vert^2) \approx 1/(1 + \sum_{\kappa }\vert Z_{\kappa} \vert^2) $ has been used. The sum in the denominator can be evaluated 
following the approach used in the phase matching section \ref{pm}. In the manuscript all results are presented for QSHHG in the band $(N-1/2) \omega_0 \le \omega_k \le (N+1/2) \omega_0$. 
Summation over this frequency band and transverse modes yields
\begin{align}
\overline{\beta} \rightarrow \beta_N \approx \frac{\beta}{1 + \vert \zeta_{N} \vert^2} \mathrm{,} \,\,\,\,\,\,\, \overline{\frak{N}} \rightarrow \frak{N}_{N} \approx 
\frac{1}{\sqrt{1 + \vert \zeta_{N} \vert^2}} \mathrm{,}
\label{wfno2mmx}
\end{align}
where $\beta_N$, $\frak{N}_{N}$ refer specifically to summing over the modes in the band $\omega_0$ about harmonic $N$. 

The multi-mode harmonic wavefunction (\ref{wfno2mm}) can be used to calculate correlation between two or more modes. Here we focus on calculating various one-mode properties of QSHHG. 
To that end we group the spatial and spectral plane-wave modes of a harmonic sideband into a single QSHH mode. 

\subsection{From plane wave modes to a single effective QSHH mode}
\label{mmto1effm}

\noindent
Experimental measurements have been made on all photons contained in single QSHH mode which is composed of many plane wave modes, as discussed above. Therefore it is advantageous to define 
operators that create or destroy photons in a QSHH mode. We define the operator \cite{rohde2007spectral}
\begin{align}
\hat{a}_N = \frac{1}{\vert \zeta_{N} \vert}\sum_{\kappa \in N} Z^*_{\kappa} \hat{a}_{\kappa}
\label{1effmop}
\end{align}
that encompasses all plane wave modes of a single quantum sideband. The operator fulfills the usual harmonic oscillator commutation relations $[\hat{a}_N,\hat{a}^{\dagger}_M] = \delta_{NM}$. 
The quadrature operators 
\begin{align}
X_{jN} = 1/(2i^{j-1}) ( \hat{a}_N + (-1)^{j-1} \hat{a}^{\dagger}_N ) \,\,\, (j=1,2) \mathrm{.}
\label{quadN}
\end{align}
also fulfill the usual bosonic commutator relations $[X_{1N}, X_{2N}] = i/2$. 

A number state of the effective harmonic mode is determined by 
\begin{align}
\vert n \rangle_{N} = \frac{1}{\sqrt{n!}} \left( \hat{a}_N^{\dagger} \right)^n \vert v \rangle_{N} \mathrm{,}
\label{1effmns}
\end{align}
where $\vert v \rangle_{N}$ is the vacuum state of the quantum sideband. Translated back into the plane wave basis, this corresponds to a sum over all combinations that have $n$ photons in the 
plane wave modes of the quantum sideband $N$. In the new basis, the wavefunction (\ref{wfno2mm}) of QSHH $N$ becomes 
\begin{align}
& \left \vert \varphi_m \right \rangle \!=\! \prod_{N} \! \frac{\frak{N}_{N}}{\sqrt{\cosh (r)}}  
\exp \! \left( \vert \zeta_{N} \vert \hat{a}_{N}^{\dagger} \hat{a}_{q}^{\dagger} \right) \!
\exp \! \left(- \beta_{N} \hat{a}_q^{\dagger 2} \right) \! \left \vert v_{N} v_q \right \rangle \! \mathrm{.}
\label{wf1effm}
\end{align}
This wavefunction will be used throughout the remaining supplement. 

We will focus on calculating the second order coherence 
\begin{align}
g^{(2)}_{N}(0) = \frac{\langle \varphi_m \vert \hat{n}_{N}^{2} - \hat{n}_{N} \vert \varphi_m \rangle}{\langle \varphi_m \vert \hat{n}_{N} \vert \varphi_m \rangle^2} \mathrm{}
\label{g2N}
\end{align}
and the quadrature variances ($j=1,2$), 
\begin{align}
& \Delta X_{jN}^2 = \langle \varphi_m \vert X^2_{jN} \vert \varphi_m \rangle - \langle \varphi_m \vert X_{jN} \vert \varphi_m \rangle^2 \nonumber \\
& = \frac{1}{4} \! \left(1 + \langle \varphi_m \vert \hat{a}_{N}^2 + \hat{a}_{N}^{\dagger 2} + (-1)^{j-1} 2 \hat{n}_{N} \vert \varphi_m \rangle \right) \mathrm{,}
\label{quadvN}
\end{align}
determined by using Eq. (\ref{quadN}) and $\langle X_{jN} \rangle = 0$.

\subsection{Probability distribution of one QSHH-mode}
\label{propqusb}

\noindent 
In this section, the QSHH photon probability distribution in a single effective mode is calculated. This applies to experiments in which the perturbative quantum field is not measured / traced out. It is sufficient to use a two-mode wavefunction $\vert \varphi_m \rangle$ consisting of the quantum perturbation and a single harmonic sideband. For the full multi-mode wavefunction, when the other harmonic sideband modes need to be traced out, the calculation proceeds in a similar manner, however the two-mode coefficients will be modified 
to higher orders in $\vert \zeta_{N} \vert^{2}$. 

The QSHHG photon number distribution is determined by $P_{N}(m) \!=\! \sum_{n} \left \vert \left \langle m n \vert \varphi_m \right \rangle \right \vert^2$ with $\vert \varphi_m \rangle$ defined 
in Eq. (\ref{wf1effm}); $m,n$ are the photon numbers of QSHH mode ($N$) and perturbative quantum mode ($q$), respectively. Using a Taylor expansion of Eq. (\ref{wf1effm}) one finds 
\begin{align}
P_{N}(m) \!=\! \frac{\frak{N}_{N}^2}{\cosh(r)} \frac{\vert \zeta_{N} \vert^{2m}}{m!} \! \sum_n \!
\frac{\vert \beta_{N}\vert^{2n}}{(n!)^2} (2n\!+\!m)! \, \mathrm{.}
\label{probh}
\end{align}
The sum in Eq. (\ref{probh}) can be evaluated by using the following relation. First we set 
$x = \vert \beta_{N} \vert$. From squeezed vacuum states \cite{gerry2023introductory} it is known that $\sum_n x^{2n} (2n)! / (n!)^2 = 1/\sqrt{1-x^2}$. 
As a result, we can write the sum in Eq. (\ref{probh}) as
\begin{align}
& \sum_n \frac{x^{2n}}{(n!)^2} (2n+m)! = \frac{d^m}{dx^m} \sum_{n} \frac{x^{2n+m}}{(n!)^2} (2n)! \nonumber \\
& = \frac{d^m}{dx^m} \frac{x^m}{\sqrt{1-x^2}} \approx \frac{(2m-1)!!}{\sqrt{1- x^2}} \left( \frac{x^{2}}{(1- x^2)} \right )^{m} + \nonumber \\
& + \frac{m}{2} (3m-1) \frac{(2m-3)!!}{\sqrt{1- x^2}} \left( \frac{x^{2}}{(1- x^2)} \right )^{m-1} + \cdots
\mathrm{.}
\label{aux1}
\end{align}
Note that we have used $r \gg 1$ so that $x \rightarrow 1$; in this limit, the leading order term comes from applying the $m$-th order derivative just to the denominator. In Eq. (\ref{aux1}) 
the leading and next order term have been included. For now we use only the leading order term; the next order is needed further below. By changing $x$ back to the original variables, the sum 
in Eq. (\ref{probh}) can be expressed as 
\begin{align}
& \sum_n \frac{\vert \beta_{N}\vert^{2n}}{(n!)^2} (2n+m)! \approx \frac{ (2m-1)!! \cosh(r) \sinh^{2m}(r)}{(1+ 2 \vert \zeta_{N} \vert^2 \cosh^2(r))^{m+1/2}} \mathrm{.}
\label{sum}
\end{align}
Inserting Eq. (\ref{sum}) in (\ref{probh}) results in 
\begin{align}
P_{N}(m) = \frac{(2m-1)!!}{(2m)!!} \frac{\left( 2 \vert \zeta_{N} \vert^2 \sinh^{2}(r) \right)^m}{(1+2 \vert \zeta_{N} \vert^2 \cosh^2(r))^{m+1/2}} \mathrm{,}
\label{probh1}
\end{align}
where $(2m)!! = 2^m m!$ has been used. 

\begin{figure}[h]
\centering
\hspace*{-0.6cm}
\includegraphics[width=9cm]{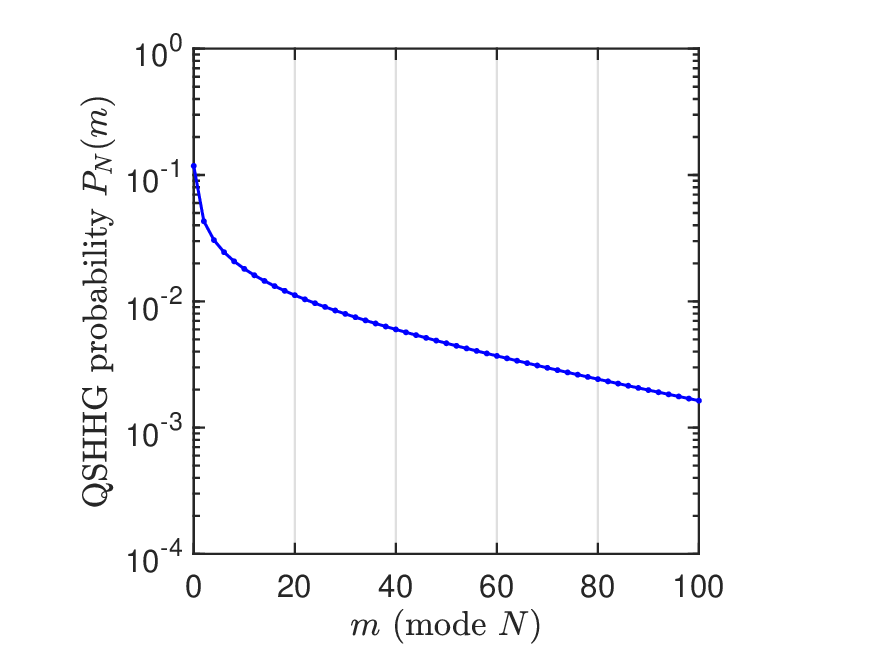}
\caption{QSHHG probability versus effective harmonic photon number $m$; exact numerical evaluation of Eq. (\ref{wf1effm}) (blue line), compared to analytical result (\ref{probh1}) (dots); 
every $m$ is populated, although dots are displayed only every second $m$ for visibility. Analytical result is valid for $r \gg 1$. Parameters: $r = 10$, $\theta=0$, and 
$\zeta_N = 5.4 \times 10^{-4}$. Smaller $r$ and larger $\zeta_N$ are used compared to the manuscript, to demonstrate validity for less intense quantum modes.} 
\label{fig1s}
\end{figure}

To test the accuracy of the analytical result, the numerical evaluation of $P_{N}(m)$ with Eq. (\ref{wf1effm}) is compared with the analytical result (\ref{probh1}) in Fig. (\ref{fig1s}). 
Excellent agreement between the numerical (full line) and the analytical (dots) result is obtained. 

Second order coherence and quadrature variance have been calculated before by using a Bogoliubov transformation \cite{lemieux2024photon}. The same results can be obtained by using wavefunction 
(\ref{wf1effm}) which is not shown here. 

\subsection{Projective measurement on quantum mode q}
\label{projq}

\noindent
We assume a measurement in which the wavefunction is projected on a number state of the quantum mode, $\vert l \rangle_q$; $m$ and $n$ again refer to general number states of the QSHH mode $N$ and perturbative quantum mode $q$. The analysis is limited to the two modes. 
The multi-mode case is discussed in the manuscript. We start from a Taylor expansion of the wavefunction (\ref{wf1effm}) limited to two modes, 
\begin{align}
\left \vert \varphi_m \right \rangle & \!=\! N_{\eta} \! \sum_{m} \! \frac{ \vert \zeta_{N} \vert^{2m}}{(2m)!} \hat{a}_{N}^{\dagger 2m} \hat{a}_{q}^{\dagger 2m} \!\! 
+ \! \frac{ \vert \zeta_{N} \vert^{2m+1}}{(2m+1)!} \hat{a}_{N}^{\dagger 2m+1} \hat{a}_{q}^{\dagger 2m+1} \nonumber \\
& \times \sum_{n} \frac{(-\beta_{N})^{n}}{n!} \hat{a}_{q}^{\dagger 2n} \left \vert v_{N} v_{q} \right \rangle \mathrm{.} 
\label{n2fix}
\end{align}
Here $N_{\eta}$ is a normalization factor to be determined. 

There are two conditions that give a fixed photon number $l$ in mode $q$: $2m+2n+\eta = l$, where $\eta = 0,1$ for even, odd $l$,
respectively. As such, $n = l_{\eta}/2-m$ with $l_{\eta} = l - \eta$ even, and the sum over $n$ can be eliminated to give 
\begin{align}
\vert \varphi_{m} \rangle & \!=\! N_{\eta} (-\beta_{N})^{l/2} \sqrt{l!} \sum_{m=0}^{l_{\eta}/2} (-1)^m 
\left( \frac{\vert \zeta_{N} \vert^{2}}{\beta_{N}} \right)^{\!\! m+\eta/2} \nonumber \\
& \times \frac{1}{\sqrt{(2m+\eta)!} \, (l_{\eta}/2-m)!} \left \vert 2m+\eta \right \rangle_{N} \vert l \rangle_q \mathrm{.}
\label{n2fix1} 
\end{align}
Projecting on a number state of the quantum mode gives the wavefunction $\vert \varphi_{N} \rangle = \mathbin{_{q} \mkern-1.5mu \langle_{}} l \! \left \vert \varphi_m(t) \right \rangle$,  
\begin{align}
\vert \varphi_{N} \rangle & \!=\! N_{\eta} (-e^{-i \theta})^{l/2} \sum_{m=0}^{l_{\eta}/2} (-1)^m 
\left( \frac{\vert \zeta_{N} \vert^{2}}{\beta_{N}} \right)^{\!\! m+\eta/2} \nonumber \\
& \times \frac{1}{\sqrt{(2m+\eta)!} \, (l_{\eta}/2-m)!} \left \vert 2m+\eta \right \rangle_{N} \mathrm{.}
\label{n2fix2} 
\end{align}
where $l$-dependent absolute value factors have been dropped, as they will disappear during normalization. 

\begin{figure}[!h]
\vspace*{-0.3cm}
\hspace*{-0.1cm}
\centering
\includegraphics[width=9.1cm]{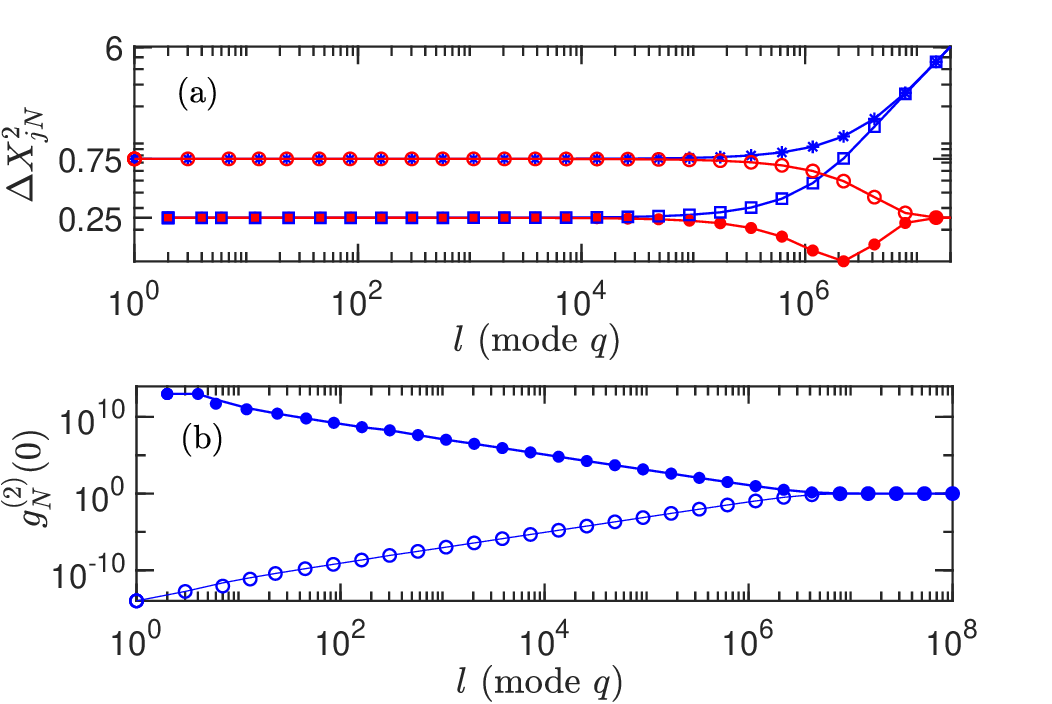} 
\caption{\label{fig2s} After a projective measurement on quantum mode q the wavefunction (\ref{n2fix2}) of the QSHH mode $N$ is obtained; with this wavefunction $\Delta X^2_{jN}$ ($j=1,2$) 
and $g^{(2)}_{N}(0)$ are evaluated and plotted in (a), (b), respectively versus photon number $l$ of the perturbative quantum mode. Parameters: $r = 10$, $\theta=0$ and $\vert \zeta_{N} \vert^2 
= 5.4\times10^{-4}$. (a), (b) exact numerical results (full lines) are compared with analytical results (markers). (a) even $l$ ($\eta = 0$, full dots and open squares) and odd $l$ ($\eta = 1$, 
open circles and stars). The blue, red plots represent $\Delta X_{1N}^2, \Delta X_{2N}^2$, respectively. (b) even ($\eta = 0$, dot), odd ($\eta = 1$, circle) photon number $l$ of mode $q$.}
\end{figure}

The normalization factor $N_{\eta}$ is determined from $\sum_m P_{N}(2m+\eta) = 1$ for $\eta = 0,1$ with 
\begin{align}
P_{N}(2m \!+\! \eta) \! = \! \frac{N_{\eta}^2 \vert \beta_{N} \vert^{\eta}}{(2m+\eta)! ((l_{\eta}/2-m)!)^2} 
\left( \frac{\vert \zeta_{N} \vert^2}{\vert \beta_{N} \vert} \right)^{2m+\eta} \mathrm{.}
\label{Pm} 
\end{align}
To proceed with the analytical derivation, the approximation 
\begin{align}
\frac{1}{(l_{\eta}/2-m)!} \approx \frac{1}{(l_{\eta}/2)!} (l_{\eta}/2)^m \left(1 - \frac{m(m+1)}{l_{\eta}} \right) \mathrm{}
\label{approx} 
\end{align}
is used. For intense quantum fields $l_{\eta}$ ranges to $ > 10^{12}$ and $m \sim 100-1000$. As $m/l_{\eta} \ll 1$, the zero order approximation in $m/l_{\eta}$ is sufficient, 
and the second term in the bracket of Eq. (\ref{approx}) is neglected. Although this approximation is not expected to hold for very small $m$, it still gives decent results, as 
will be demonstrated in Fig. \ref{fig2s}. Further, for the same reason the sum over $m$ can be extended to infinity in the large $l_{\eta}$ limit. The approximate wavefunction is obtained as
\begin{align}
& \vert \varphi_{N} \rangle \!=\! N_{\eta} (-e^{-i \theta})^{l/2} \sum_{m=0}^{\infty} (-1)^m 
\frac{\sqrt{\alpha_{N}}^{2m+\eta}}{\sqrt{(2m+\eta)!}}  \left \vert 2m+\eta \right \rangle_{N} \nonumber \\
& N_0 \approx \frac{1}{\sqrt{\cosh(\vert \alpha_{N} \vert)}} \,\,\,\,\,\,\, \alpha_{N} = \frac{l_{\eta} \vert \zeta_{N} \vert^2}{2 \beta_{N}} \nonumber \\
& N_1 \approx \frac{1}{\sqrt{\sinh(\vert \alpha_{N} \vert)}}
\label{n2fixN} 
\end{align}
To leading order, $N_{\eta}$ is determined by using the relations 
\begin{align}
& \sum_{m} \frac{\vert \alpha_{N} \vert^{2m}}{(2m)!} = \cosh(\vert \alpha_{N} \vert) \nonumber \\ 
& \sum_{m} \frac{\vert \alpha_{N} \vert^{2m+1}}{(2m+1)!} = \sinh(\vert \alpha_{N} \vert) \mathrm{.}
\nonumber 
\end{align}

Using the projected wavefunction (\ref{n2fixN}) and the same approximations as above, we find 
\begin{align}
& g^{(2)}_{N} \approx \frac{1}{\tanh^2(\vert \alpha_{N} \vert)} \,\,\,\,\,\,\,\, \mathrm{for} \, \eta = 0 \nonumber \\
& g^{(2)}_{N} \approx \tanh^2(\vert \alpha_{N} \vert) \,\,\,\,\,\,\,\,\, \mathrm{for} \, \eta = 1 \mathrm{}
\label{n2fixg2} 
\end{align}
and for the quadratures ($j=1,2$)
\begin{align}
& \Delta X_{j N}^2 \!\approx\! \frac{1}{4} \left[1 \!+\! 2 \left(\vert \alpha_{N} \vert \tanh(\vert \alpha_{N} \vert) \!+\! (-1)^j \mathrm{Re}[\alpha_{N}] \right) \right] \! 
\mathrm{,} \,\, \eta = 0 \nonumber \\
& \Delta X_{j N}^2 \!\approx\! \frac{1}{4} \left[ 1 \!+\! 2 \left( \frac{\vert \alpha_{N} \vert}{\tanh(\vert \alpha_{N} \vert)} \!+\! 
(-1)^j \mathrm{Re}[\alpha_{N}] \right) \right] \! \mathrm{,} \,\,\,\,\,\, \eta = 1 \mathrm{.}
\label{n2fixquad} 
\end{align} 

In Fig. (\ref{fig2s}) the exact numerical results for $g^{(2)}_{N}(0)$ and $\Delta X^2_{jN}$ (full lines), obtained from the wavefunction (\ref{wf1effm}), are compared with the
analytical results (\ref{n2fixg2}) and (\ref{n2fixquad}) (markers). The agreement is excellent, validating the analytical derivation. 

\begin{figure}[!h]
\centering
\hspace*{-0.8cm}
\includegraphics[width=9.3cm]{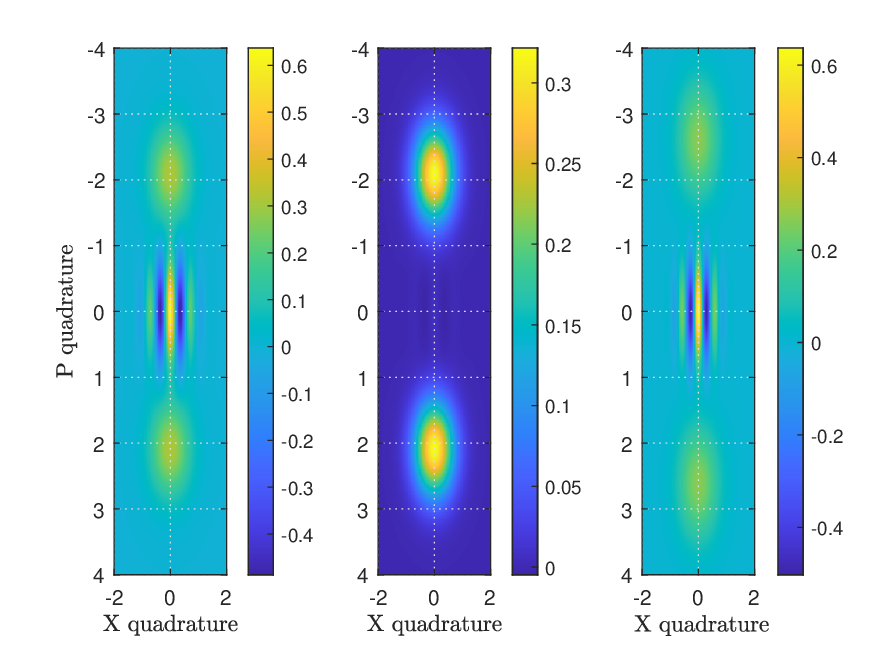} 
\caption{\label{fig3s} The Wigner function of the wavefunction (\ref{n2fixN}) is plotted versus $X,P$ which are the quadratures. Parameters are taken from Fig. 1(c) of the manuscript for harmonic sideband $N=8$ and $\theta = 0$; $r=13.6$ and $\langle \hat{n} \rangle_N = 67.2$; projected, perturbative quantum photon number $l=10^{10}$; photon number resolution $\Delta l=0, 100, 5 \times 10^{9}$ for (a), (b), (c), respectively; in (c) parity measurement is assumed so that the Wigner function is only averaged over 
even states. }
\end{figure}

In Fig. \ref{fig3s}(a)-(c) the Wigner function of the projected wavefunction (\ref{n2fixN}) is plotted for quantum photon number 
$l=10^{10}$. The parameters are taken from Fig. 1(c) of the manuscript for harmonic sideband $N=8$ and $\theta = 0$; $r=13.6$ and 
$\langle \hat{n} \rangle_N = 67.2$. Here we study the effect of limited photon number resolution. It is extremely challenging to 
resolve such high photon numbers down to a single photon. The uncertainty of photon number resolution is referred to as $\Delta l$. 
A clear modulation of the Wigner function with negative and positive parts can be seen in Fig. \ref{fig3s}(a) for a projective 
measurement with single-photon resolution, $\Delta l=0$. The modulation is completely averaged out in Fig. \ref{fig3s}(b) for a 
photon resolution $\Delta l = 100$. This is due to the phase oscillation of $\pi$ between even and odd $l$ states. When the parity 
is known, i.e. only even or odd states are measured, then the non-classical structure remains for the most part preserved, even for 
poor photon number resolution, see Fig. \ref{fig3s}(c) for $\Delta l = 5\times 10^{9}$.

\subsection{Projective measurement on harmonic modes $N$}
\label{projN}

\noindent
In a projective measurement $m$ photons in the QSHH mode $N$ are obtained. The resulting wavefunction of quantum mode $q$, $\mathbin{_{N} \mkern-1.5mu \langle_{}} m \! \left \vert 
\varphi_m \right \rangle = \vert \varphi_q \rangle $, is obtained from Eq. (\ref{wf1effm}) limited to two modes. Taylor expansion of the wavefunction followed by the projection on the QSHH number state gives 
\begin{align}
\left \vert \varphi_q \right \rangle \!=\! N_m \sum_{n=0}^{\infty} (-1)^n \frac{\sqrt{(2n+m)!}}{n!} \beta_{N}^n \left \vert 2n+m \right \rangle_{q} \mathrm{,}
\label{n1fix} 
\end{align}
where factors that do not depend on $n$ have been dropped, as they will fall away during normalization anyway. The normalization factor is determined by 
\begin{align}
& \left \langle \varphi_q \vert \varphi_q \right \rangle \!=\! \sum_{n} P_{q}(n) = N_m^2 B_m = 1 \nonumber \\
& B_m = \sum_{n=0}^{\infty} \vert \beta_{N} \vert^{2n} \frac{(2n+m)!}{(n!)^2} \mathrm{,}
\label{n1fixnrm} 
\end{align}
which has the same form as Eq. (\ref{probh}) and thus, can be evaluated with the same method as in Eq. (\ref{sum}). This gives 
\begin{align}
B_m \! & \approx \! \frac{(2m\!-\!1)!! \cosh(r)}{\sqrt{1 \!+\! 2 \vert \zeta_{N} \vert^2 \cosh^2(r)}} \biggl[ \! \left( \! \frac{\sinh(r)}{1 \!+\! 2 \vert \zeta_{N} \vert^2 \cosh^2(r)} 
\! \right)^{\!\! m} \nonumber \\
& + \frac{m}{2} \frac{3m-1}{2m\!-\!1} \left( \frac{\sinh(r)}{1 + 2 \vert \zeta_{N} \vert^2 \cosh^2(r)} \right)^{m-1} \biggr]
\label{n1fixnrm} 
\end{align}

For $g^{(2)}_{N}(0)$ we need the expectation values of $\hat{n}_q$ and $\hat{n}_q^2$; For the calculation it is sufficient to limit Eq. (\ref{n1fixnrm}) to the leading order term; we have
\begin{align}
\left \langle \varphi_q \left \vert \hat{n}_q \right \vert \varphi_q \right \rangle \!=\! N_m^2 \sum_{n=0}^{\infty} (-1)^n \frac{(2n+m)!}{(n!)^2} \beta_{N}^n (2n+m)  
\mathrm{.}
\label{n1fix1} 
\end{align}
By rewriting the last term $2n+m = (2n+m+1) - 1$ we can use the same procedure as was used above for the normalization, and obtain 
\begin{align}
\left \langle \varphi_q \left \vert \hat{n}_q \right \vert \varphi_q \right \rangle \!=\! N_m^2 B_{m+1} \!-\! 1 \approx 
(2m+1) \frac{\sinh^2(r)}{1 \!+\! 2 \vert \zeta_{N} \vert^2 \cosh^2(r)} \mathrm{.}
\label{n1fix2} 
\end{align}
Further, 
\begin{align}
\left \langle \varphi_q \left \vert \hat{n}_q^2 \right \vert \varphi_q \right \rangle \!=\! N_m^2 \sum_{n=0}^{\infty} (-1)^n \frac{(2n+m)!}{(n!)^2} \beta_{N}^n (2n+m)^2  
\mathrm{.}
\label{n1fixn22} 
\end{align}
Again, the last term is reexpressed as $(2n+m)^2 = (2n+m+2)(2n+m+1) -3(2n+m+1) +1 \approx (2n+m+2)(2n+m+1)$ which gives the leading order term in the $r \gg 1$ limit, $N_m^2 B_{m+2}$; this results 
in 
\begin{align}
\left \langle \varphi_q \left \vert \hat{n}_q^2 \right \vert \varphi_q \right \rangle \! \approx \!
(2m+3) (2m+1) \left( \frac{\sinh^2(r)}{1 + 2 \vert \zeta_{N} \vert^2 \cosh^2(r)} \right)^2 \! \mathrm{.}
\label{n1fixn22x} 
\end{align}

Equations (\ref{n1fix2}) and (\ref{n1fixn22x}) determine $g^{(2)}_{N}(0)$ as
\begin{align}
g^{(2)}_{N}(0) \approx \frac{(2m+3)(2m+1) \sinh^4(r)}{(2m+1)^2 \sinh^4(r)} = 1 + \frac{2}{2m+1} \mathrm{.}
\label{n1fixg2} 
\end{align}

For the quadratures we still need to evaluate 
\begin{align}
& \left \langle \varphi_q \left \vert \hat{a}_q^{\dagger 2} \right \vert \varphi_q \right \rangle \!\approx\! \frac{-1}{2 \beta_{\kappa}} \left( N_m^2 B_{m+1} - (m+1) \right) \nonumber \\
& \left \langle \varphi_q \left \vert \hat{a}_q^{2} \right \vert \varphi_q \right \rangle_m \!\approx\! \frac{-1}{2 \beta_{\kappa}^*} \left( N_m^2 B_{m+1} - (m+1) \right)
\mathrm{.}
\label{n1fixa2p2} 
\end{align}
Inserting Eqs. (\ref{n1fixa2p2}) and (\ref{n1fix2}) in Eq. (\ref{quadvN}) gives ($j=1,2$)
\begin{align}
\Delta X_{jN}^2 \! & =\! \frac{1}{4} \biggl( 2 N^2_m B_{m+1} - 1 + (-1)^j \frac{1 + \vert \zeta_{N} \vert^2}{\tanh(r)} \cos(\theta) \nonumber \\
& \times \left( 2 N_m^2 B_{m+1} - 2(m+1) \right) \biggl) \mathrm{.}
\label{n1fixDxy} 
\end{align}
In order to evaluate Eq. (\ref{n1fixDxy}) correctly, the next order term in Eq. (\ref{n1fixnrm}) has to be carried along as well. We use $\vert \zeta_{N} \vert^2 \cosh^2(r) \gg 1$ and keep 
$\vert \zeta_{N} \vert^2 \ll 1$ only to leading order. This results in ($j=1,2$)
\begin{align}
& \Delta X^2_{jN} \! \approx \! \frac{ A_j(r, \theta)\!+\! \vert \zeta_{N} \vert^2 (1 \!+\! \vert \zeta_{N} \vert^2 \cosh^2(r))}{1+ 2 \vert \zeta_{N} \vert^2 \cosh^2(r)} \frac{(2m+1)}{4}
\nonumber \\
& + \frac{m(m-1)}{2(2m-1)} \left(1 +(-1)^j \cos(\theta) \frac{1 + \vert \zeta_{N} \vert^2}{\tanh(r)} \right) \mathrm{.}
\label{n1fixquad} 
\end{align}
with 
\begin{align}
A_j(r,\theta) & \!=\! \cosh^2(r) \!+\! \sinh^2(r) \nonumber \\
& + 2 (-1)^j \! \cosh(r) \sinh(r) \cos(\theta) \mathrm{.}
\nonumber
\end{align}
For the case $\theta = 0$ one obtains
\begin{align}
\Delta{X_{1N}^2} \! & \approx \! \frac{2m+1}{4}\frac{ (\cosh(r) \!-\! \sinh(r))^2}{1+ 2 \vert \zeta_{N} \vert^2 \cosh^2(r)} \nonumber \\
& + \frac{\vert \zeta_{N} \vert^2}{4} \left(1 + \frac{1}{2(2m-1)} \right) \mathrm{.}
\label{n1fixquadx} 
\end{align}
The first term is proportional to the quadrature variance of squeezed vacuum. The additional pre-factors in the first 
term and the second term come from the mixing between harmonic modes and perturbative quantum mode. 

\begin{figure}[!h]
\hspace*{-0.3cm}
\centering
\includegraphics[width=9.5cm]{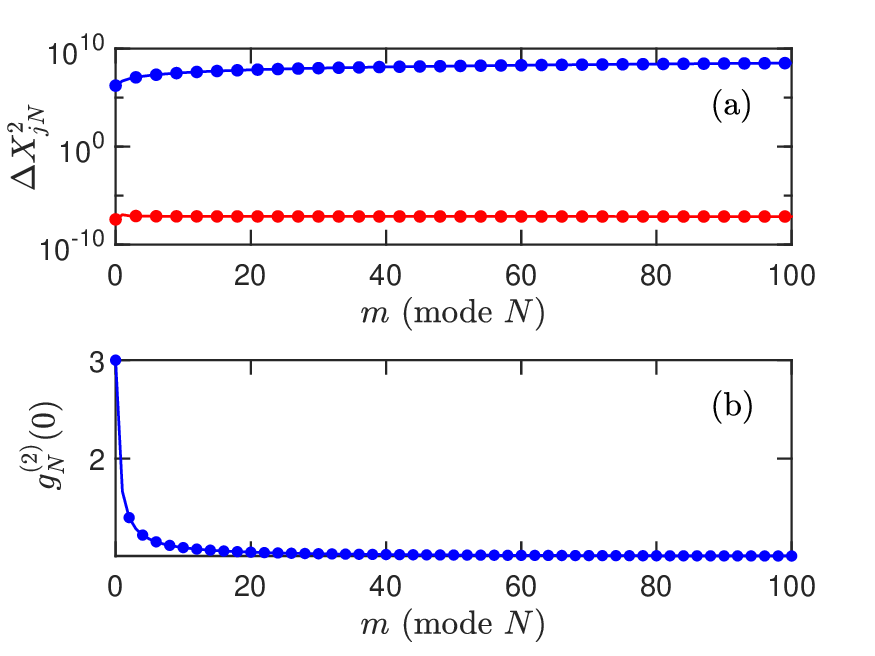} 
\caption{\label{fig4s} In a projective measurement $m$ photons are measured in a QSHH mode. The resulting wavefunction depends on the perturbative quantum mode alone. Variance of 
quadratures ($j=1$, blue), ($j=2$, red) (a), and $g^{(2)}_{N}(0)$ of quantum mode $q$ (b) versus QSHH photon number $m$ are plotted. Parameters: $r = 10$, $\theta=0$, and $\zeta_N = 5.4\times10^{-4}$. 
Numerical results from evaluation of Eqs. (\ref{wf1effm}) (full lines) are compared with analytical results (dots).}
\end{figure}

In Fig. (\ref{fig4s}) quadrature variances (a) and second order correlation function (b) of quantum mode $q$ are shown versus photon number $m$ in QSHH mode $N$. Numerical (full lines) and 
analytical results (dots) match well. 

\begin{figure}[h]
\centering
\hspace*{-0.6cm}
\includegraphics[width=9.6cm]{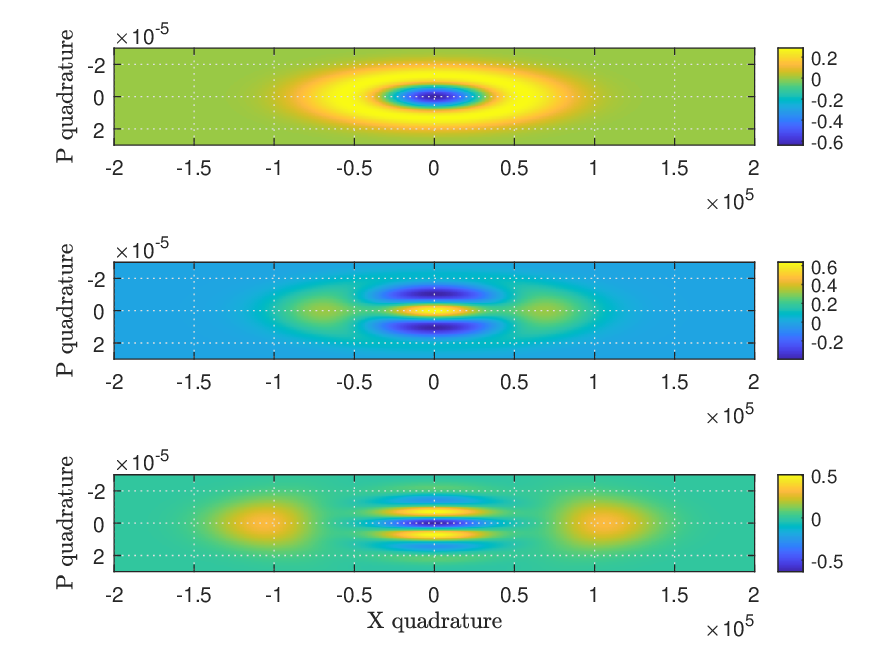} 
\caption{\label{figfinal33} The Wigner functions of Eq. (\ref{n2fixN}) plotted versus $X,P$, 
the quadratures, for (a) $m=1$,  (b) $m=2$ and (c) m=5, with $r=13.6$ and $\theta=0$.}
\end{figure}

Equation (\ref{n1fix}) presents an m-photon added squeezed vacuum state \cite{kun2010nonclassicality}. Its Wigner function is 
plotted in Figure \ref{figfinal33} for m=1 (a), m=2 (b) and m=5 (c); parameters are r=13.6 and $\theta=0$. 
The Wigner function becomes negative, a clear indication of nonclassicality, and features sub-shot noise oscillations in the squeezed
quadrature. Also note the strongly elongated shape of the Wigner function along the $X$ quadrature. 
 
\newpage
\bibliographystyle{apsrev4-2} 
\bibliography{supplement} %

\clearpage